%% file: wiplus-tr.tex
\documentclass[10pt,journal,compsoc, onecolumn]{IEEEtran}
 \pdfoutput=1
\usepackage{etex}
\usepackage{graphicx}
\usepackage[utf8]{inputenc}

\usepackage{etoolbox}

\usepackage[procnames]{listings}
\usepackage{color}

\usepackage{amssymb}
\graphicspath{{figures/}}

\usepackage{url}
\usepackage{wrapfig}
\usepackage{units}

\usepackage[printonlyused]{acronym}

\usepackage{calc}
\usepackage{amsmath}
\usepackage{amssymb}
\usepackage{amsfonts}

\usepackage{algorithm}
\usepackage{algpseudocode}
\usepackage{color}
\usepackage[table]{xcolor} 

\usepackage[font=footnotesize]{caption}

\usepackage{subcaption}
\usepackage{stfloats}
\usepackage{nicefrac}

\usepackage{booktabs}
\newcolumntype{C}[1]{>{\centering\arraybackslash}p{#1}}

\usepackage{color}

\newcolumntype{C}[1]{>{\centering\arraybackslash}p{#1}}

\usepackage{booktabs}

\definecolor{BgGray}{gray}{0.7}%
\definecolor{BgGray2}{gray}{0.96}%
\definecolor{RowColorOdd}{named}{BgGray2}%
\definecolor{RowColorEven}{named}{white}%
\definecolor{comments}{gray}{.5}
\definecolor{Gray}{gray}{0.85}

\definecolor{keywords}{RGB}{255,0,90}
\definecolor{red}{RGB}{160,0,0}
\definecolor{green}{RGB}{0,150,0}
\definecolor{deepblue}{rgb}{0,0,0.5}
\definecolor{deepred}{rgb}{0.6,0,0}
\definecolor{deepgreen}{rgb}{0,0.5,0}

\usepackage{multirow}

\usepackage[utf8]{inputenc}
\usepackage{tabularx}
\usepackage{multicol}
\usepackage{graphicx}

\usepackage{tikz}

\usepackage[headsepline,plainheadsepline,footsepline,plainfootsepline]{scrpage2}
\usepackage{mathptmx}
\usepackage[scaled=.92]{helvet}
\usepackage{courier}
\usepackage[multiuser,nomargin,marginclue,footnote]{fixme}
\usepackage{upgreek}

\usepackage{calc}
\usepackage{amsmath}
\usepackage{amssymb}
\usepackage{amsfonts}

\definecolor{BgGray}{gray}{0.1}%
\definecolor{BgGray2}{gray}{0.96}%
\definecolor{RowColorOdd}{named}{BgGray2}%
\definecolor{RowColorEven}{named}{white}%
\definecolor{comments}{gray}{.5}
\definecolor{Gray}{gray}{0.85}

\newcommand\colheadbegin{\hline\rowcolor{BgGray}}
\newcommand\colheadend{\\\hline}

\usepackage{pifont}
\usepackage[multiuser,nomargin,marginclue,footnote]{fixme}
\fxusetheme{color}
\fxsetup{targetlayout=colorcb}
\FXRegisterAuthor{all}{anall}{ALL}
\FXRegisterAuthor{tolja}{antolja}{TOLJA}
\FXRegisterAuthor{mch}{anmc}{MC}
\FXRegisterAuthor{cp}{ancp}{CP}

\fxsetup{status=draft}

\definecolor{BgGray}{gray}{0.5}%

\usepackage{geometry}
\clearscrheadfoot

\ihead[\Large {\scshape TKN Technical Reports Series / Technische Universität Berlin }]{\small {TR \trnumber}}

\ifoot[{\tiny \begin{minipage}{4.0cm}Copyright at Technische Universität Berlin.\newline All Rights Reserved.\end{minipage}}]{{\tiny \begin{minipage}{4.0cm}Copyright at Technische Universität Berlin.\newline All Rights Reserved.\end{minipage}}}
\ofoot[Page \pagemark]{Page \pagemark}

\newcommand{\trnumber}{TKN-17-0001}
\newcommand{\trdate}{February 2017}
\newcommand{\trauthor}{Michael Olbrich, Anatolij Zubow, Sven Zehl and Adam Wolisz}
\newcommand{\tremail}{\{olbrich, zubow, zehl, wolisz\}@tkn.tu-berlin.de}
\newcommand{\trtitle}{Towards LTE-U Interference Detection, Assessment and Mitigation in 802.11 Networks using Commodity Hardware}

\ifCLASSOPTIONcompsoc 
  \usepackage[nocompress]{cite}
\else
  \usepackage{cite}
\fi

\ifCLASSINFOpdf
\else
\fi

\begin{document}

\providetoggle{techreport}
\settoggle{techreport}{true}

\input{packages/tr_cover.tex}

\title{}
\author{}



\newpage
\input{sections/abstract}

\newpage
\input{sections/introduction}

\input{sections/primer}

\input{sections/understanding}

\input{sections/problem}

\input{sections/approach}


\input{sections/apps}

\input{sections/eval}

\input{sections/simulation}

\input{sections/related}

\input{sections/conclusion}

\bibliographystyle{IEEEtran}
\bibliography{IEEEabrv,biblio}

\end{document}

%% file: packages/tr_cover.tex



{
\sffamily

\thispagestyle{empty}

\begin{tabularx}{\columnwidth}{cXc}
  \includegraphics[height=1cm]{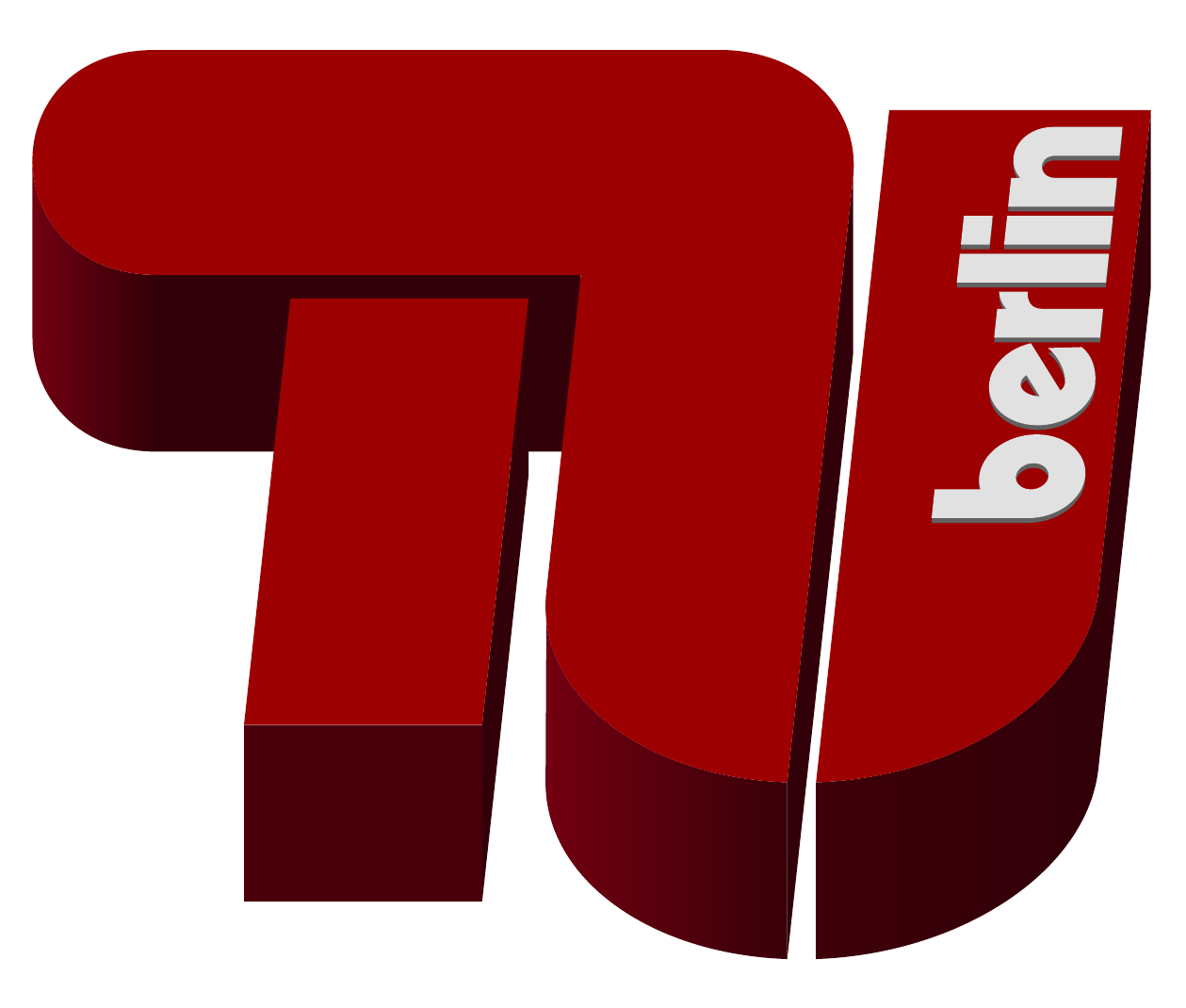}
  & &
  \includegraphics[height=1cm]{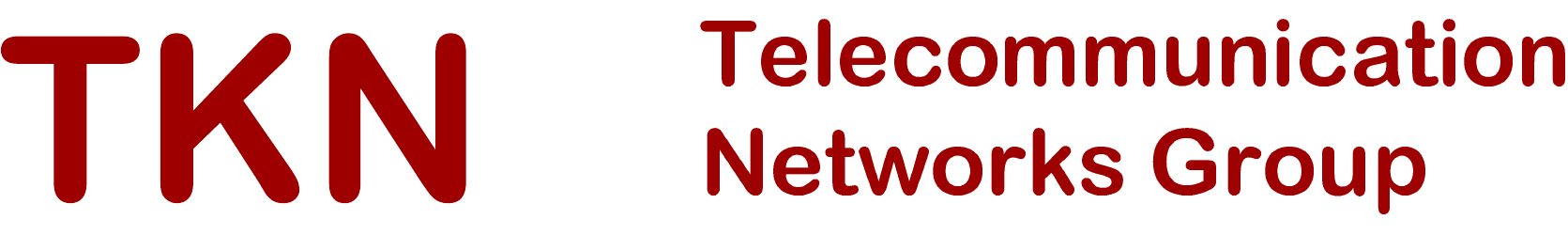}
  \\
\end{tabularx}

\vspace{1.0cm}

\begin{center}
{\huge
\noindent
Technische Universität Berlin

\vspace{0.5cm}

\noindent
Telecommunication Networks Group

\begin{center}
\rule{15.5cm}{0.4pt}
\end{center}
}
\end{center}

\begin{minipage}[][11.0cm][c]{14.5cm}
{\Huge

\begin{center}
\trtitle
\end{center}

\begin{center}
{\LARGE \trauthor} \\
{\Large \tremail}
\end{center}

\begin{center}
Berlin, \trdate
\end{center}

\vspace{0.5cm}

}

\begin{center}
\setlength{\fboxrule}{2pt}\setlength{\fboxsep}{2mm}
\fbox{TKN Technical Report \trnumber}
\end{center}

\end{minipage}

\setlength{\fboxrule}{0.4pt}
\setlength{\fboxsep}{0.4pt}

\begin{center}

  \rule{15.5cm}{0.4pt}

  \vspace{0.5cm}

  {\huge {TKN Technical Reports Series}}

  \vspace{0.5cm}

  {\huge Editor: Prof. Dr.-Ing. Adam Wolisz}

  \vspace{0.5cm}

 \end{center}

}

%% file: sections/abstract.tex
\begin{abstract}
\iftoggle{techreport}{
\begin{center}
\begin{minipage}[t]{0.7\textwidth}
}%
{
}
We propose WiPLUS -- a system that enables WiFi to deal with the stealthy invasion of LTE-U into the frequency bands used by WiFi. Using solely MAC layer information extracted passively, during runtime, out of the hardware registers of the WiFi NIC at the WiFi access point, WiPLUS is able to:  i) detect interfering LTE-U signals, ii) compute their duty-cycles, and iii) derive the effective medium airtime available for each WiFi link in a WiFi Basic Service Set (BSS).
Moreover WiPLUS provides accurate timing information about the detected LTE-U ON and OFF phases enabling advanced interference mitigation strategies such as interference-aware scheduling of packet transmissions, rate adaptation and adaptive channel bonding.

WiPLUS does not require any modifications to the WiFi client stations and works with commodity WiFi APs where it has a simple software installation process.

We present the design, the implementation details and the evaluation of the WiPLUS approach.
Evaluation results reveal that it is able to accurately estimate the effective available medium airtime for each link in a WiFi BSS under a wide range of LTE-U signal strengths with a root-mean-square error of less than 3\,\% for the downlink and less 10\,\% for the uplink.
\iftoggle{techreport}{
\end{minipage}
\end{center}
}%
{
}
\end{abstract}

%% file: sections/introduction.tex
%
%
\section{Introduction}

Cellular network operators like Verizon~\cite{verizonLteU} are starting to offload data traffic in unlicensed 5\,GHz ISM spectrum using LTE Unlicensed (LTE-U~\cite{lteu_forum}). However, this part of the radio spectrum is also used by existing and future IEEE 802.11 standards, e.g. 802.11a/ac/ax. 

Multiple studies have been carried out to identify the effects of LTE-U on WiFi~\cite{ho2017u}. In particular, Jindal et al.~\cite{lteu_study_google} showed, as LTE-U duty-cycling does not implement listen before talk mechanisms (LBT), it can under some conditions even disproportionately reduce WiFi throughput performance. Moreover, interference from LTE-U with moderate power can be even more harmful to WiFi than high-power interference.

\iftoggle{techreport}{
\begin{wrapfigure}[17]{L}{0.7\textwidth}
	\centering
	\includegraphics[width=0.7\textwidth]{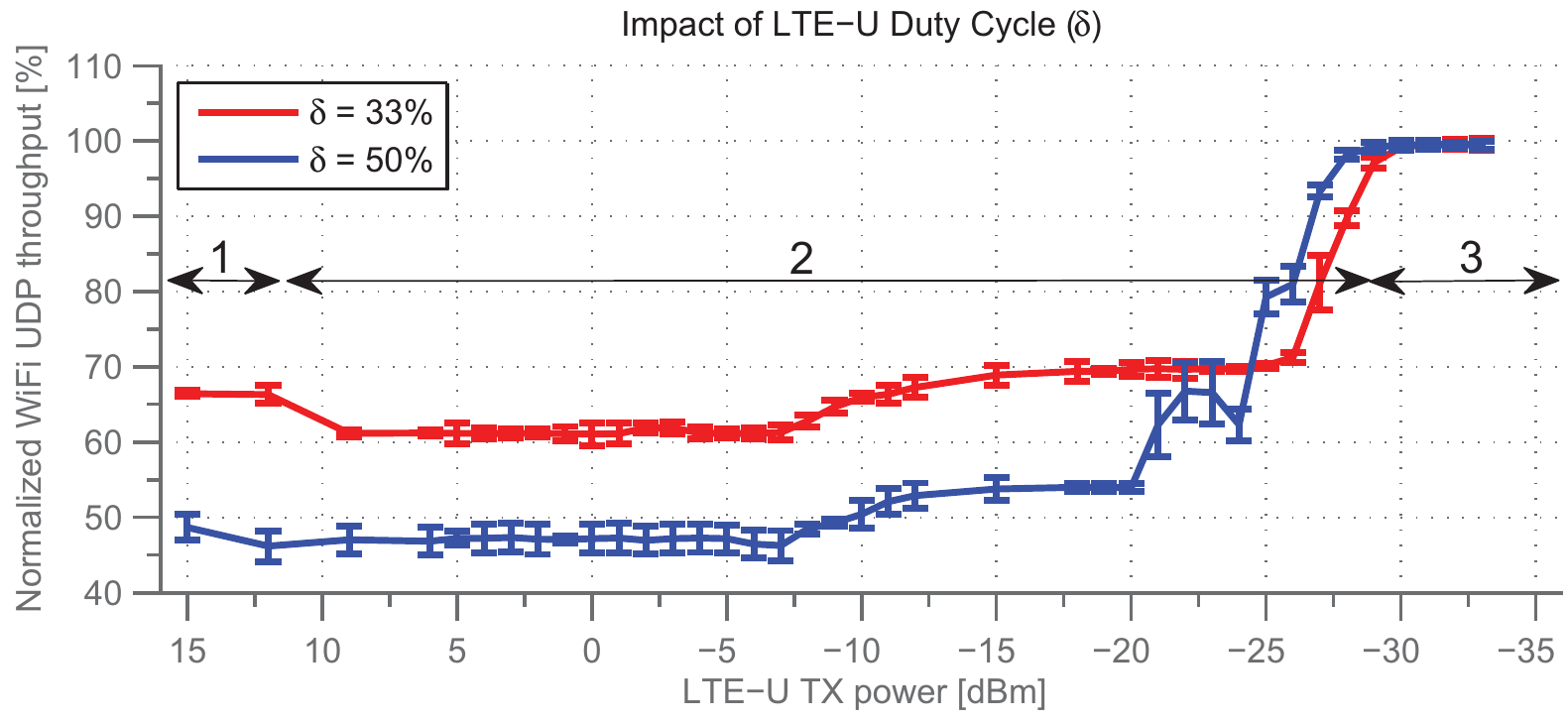}	
	\caption{Degradation in UDP throughput of high quality WiFi link in presence of a LTE-U BS operating in the same Rf band and for various LTE-U transmit power values. Plot shows mean and standard deviation. See Sec.~\ref{sec:exp_sec} for more details.}
	\label{fig:udp_degradation}
\end{wrapfigure}	
}%
{
\begin{figure}[h!]
	\centering
\includegraphics[width=1\linewidth]{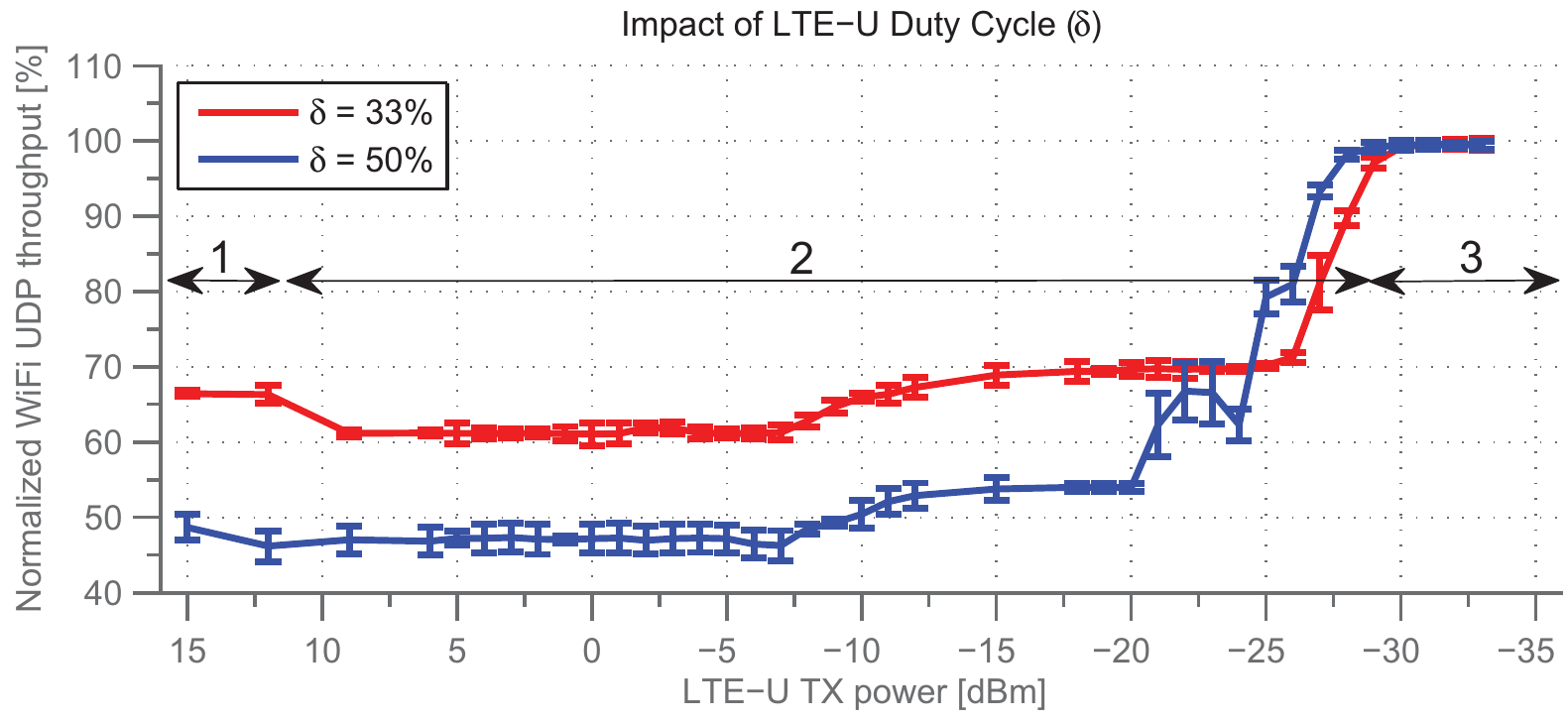}	
	\caption{Degradation in UDP throughput of high quality WiFi link in presence of LTE-U BS operating in the same Rf band and for various LTE-U TX power values. Plot shows mean and standard deviation. See Sec.~\ref{sec:exp_sec} for more details.}
	\label{fig:udp_degradation}
\end{figure}	
}

Fig.~\ref{fig:udp_degradation} shows results of our own experiments in which a single high quality WiFi link was suffering from interference of a LTE-U Base Station (BS). In particular the normalized UDP throughput of the WiFi link under LTE-U interference, relative to the non-interfered WiFi link, which corresponds to the effective available medium airtime, as a function of different interfering signal strengths and LTE-U duty-cycles, is presented. We can clearly see the impact of the LTE-U duty cycle on the WiFi performance even at very low power levels, i.e. distant LTE-U BS.


To be able to cope with this impact and its effects, an approach that first, enables Wi-Fi to detect the LTE-U interference and second, enables to quantify the effective available medium airtime of each link (downlink and uplink) during runtime, is needed. Afterwards, this knowledge enables to apply rapid interference management techniques such as
%
%
assignment of radio channels, or for network load balancing reasons, i.e. optimizing the client station (STA) associations across WiFi APs. Furthermore, information about the exact timings of the LTE-U ON and OFF phases allows to apply techniques like interference-aware packet scheduling, i.e. transmission during LTE-U OFF phases, and adaptive channel bonding, i.e. using secondary channel during LTE-U OFF phases only.

\medskip

\noindent \textbf{Contributions: } WiPLUS is a method to estimate the effective available medium airtime on each downlink (DL) and uplink (UL)  in an 802.11 BSS under LTE-U interference through passive determination of LTE-U BS’s duty-cycle and its timing in real-time at WiFi APs. WiPLUS has a simple software installation process at the WiFi AP and does not need any modification within the WiFi STAs. WiPLUS was prototypically implemented using commodity 802.11 hardware. Results from experimental evaluation revealed that WiPLUS is able to accurately estimate the effective available air-time in the DL at a wide range of LTE-U signal strengths. Moreover, evaluations from system-level simulations revealed that in order to accurately predict the uplink, MAC acknowledgment frames need to be rate controlled.


%% file: sections/primer.tex
%
%
\section{LTE-U Primer}

\iftoggle{techreport}{
	\begin{wrapfigure}[7]{R}{0.55\textwidth}
		\centering
		\includegraphics[width=0.5\textwidth]{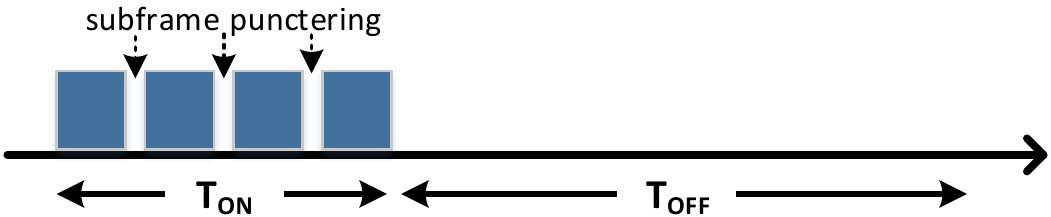}
		\caption{Duty cycling in LTE-U.}
		\label{fig:lte-u-overview}
	\end{wrapfigure}	
}%
{
}

LTE-U being specified by LTE-U forum~\cite{lteu_forum} is the first technology to be deployed in which the LTE lower layers directly use the unlicensed band, i.e. aggregation on modem level.
Here, LTE-U makes use of the LTE carrier aggregation framework by utilizing the unlicensed band as a secondary cell in addition to the licensed anchor that will serve as the primary cell~\cite{ni_lte_whitepaper}. The current LTE-U specification is restricted to DL traffic. The LTE-U channel bandwidth is set to 20\,MHz which corresponds to the smallest channel width in WiFi. 

Fig.~\ref{fig:lte-u-overview} illustrates the duty cycled unlicensed channel access of LTE-U. LTE-U considers mechanisms enabling co-existence with WiFi. Therefore, LTE-U BSs actively observe the channel for WiFi transmissions to estimate channel activity for dynamic channel selection and adaptive duty cycling. A mechanism called carrier sense adaptive transmission (CSAT) is used to adapt the duty cycle~\cite{lteu_qualcom2016}, i.e. by modifying the $T_{\mathrm{ON}}$ and $T_{\mathrm{OFF}}$ values, to achieve fair sharing. Moreover, LTE-U transmissions contain frequent gaps in the ON-cycle, which allow WiFi to transmit delay-sensitive data. Note that LBT is not applied in LTE-U before transmission of packets in the ON-cycle. WiFi on the other hand cannot decode LTE-U frames, but has to rely on energy-based carrier sensing.

\iftoggle{techreport}{
}%
{
\begin{figure}[h!]
	\centering
	\includegraphics[width=0.7\linewidth]{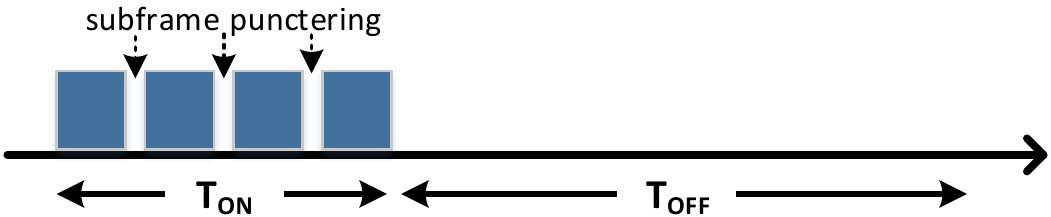}
	\caption{Duty cycling in LTE-U.}
	\label{fig:lte-u-overview}
\end{figure}	
}


%% file: sections/understanding.tex
%
%
\section{Understanding the Impact of LTE-U Co-Channel Interference on WiFi}
\label{sec:under}

Our system model is shown in Fig.~\ref{fig:system_model}. Here we have a WiFi BSS consisting of a single AP serving multiple STAs. In addition there is a LTE-U BS serving several LTE User Equipment nodes (UE). The LTE-U BS is operating in the DL using the same unlicensed radio spectrum as the WiFi BSS. 
We can further extend our system model and incorporate multiple co-located LTE-U cells, even from different operators, as long as they have identical and (time) aligned duty-cycles like proposed by Cano et al.~\cite{cano2015}.

\iftoggle{techreport}{
	\begin{wrapfigure}[13]{R}{0.55\textwidth}
		\centering
		\includegraphics[width=0.9\linewidth]{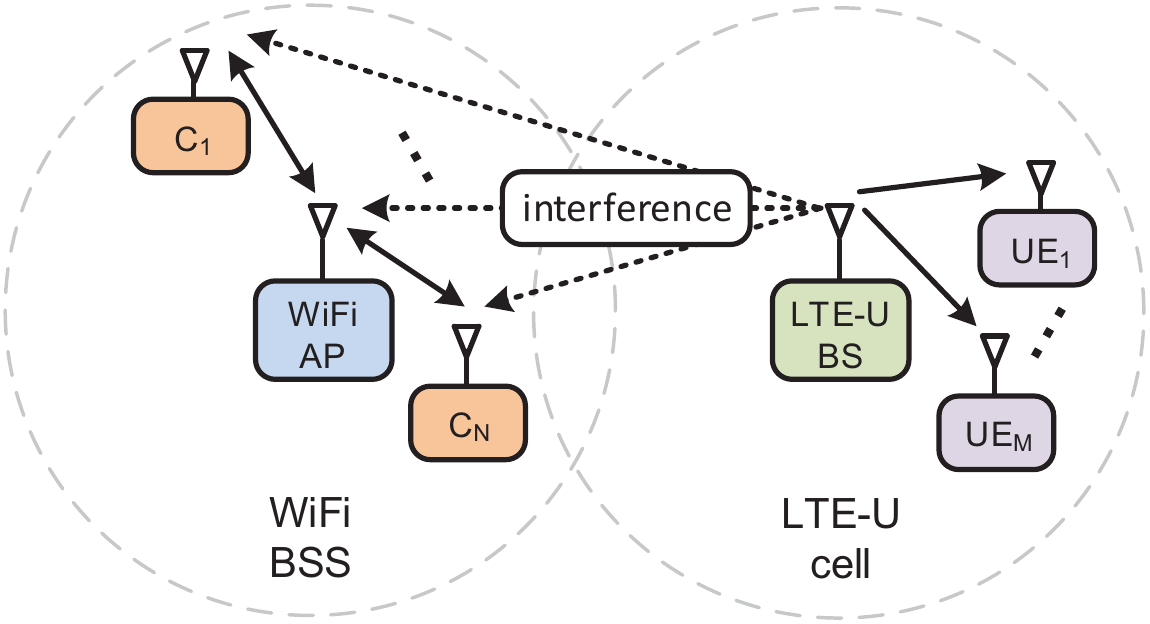}
		\caption{System model --- WiFi BSS co-located with LTE-U cell. The LTE-U DL traffic creates interference on WiFi BSS DL and UL traffic.}
		\label{fig:system_model}
	\end{wrapfigure}	
}%
{
	\begin{figure}[h!]
		\centering
		\includegraphics[width=0.7\linewidth]{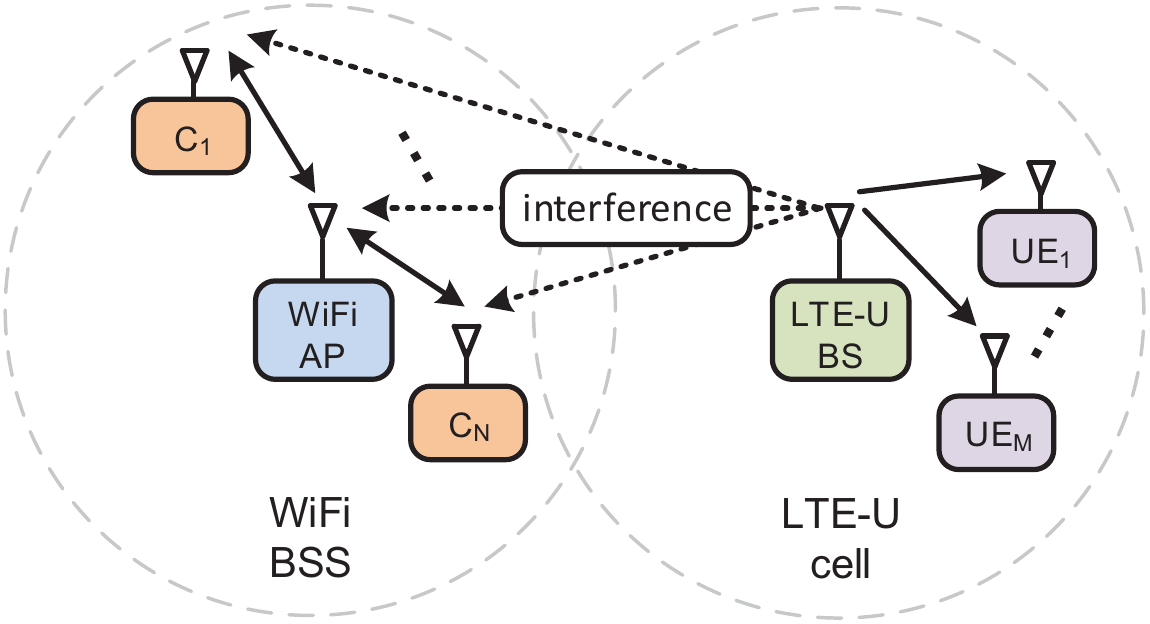}
		\caption{System model --- WiFi BSS co-located with LTE-U cell. The LTE-U DL traffic creates interference on WiFi BSS DL and UL traffic.}
		\label{fig:system_model}
	\end{figure}	
}

The LTE-U DL signal may or may not impact the WiFi communication in three ways, namely,
\begin{enumerate}
	\item \textit{Blocking medium access} by triggering the Energy Detection (ED) physical Carrier Sense (CS) mechanism of WiFi or
	\item \textit{Corrupting packets} due to co-channel interference from LTE-U.
	\item \textit{No impact} due to insignificant co-channel interference from LTE-U.
\end{enumerate}

Whether the first or the second has an impact depends on the received LTE-U signal strength at the WiFi transmitter and receiver. At high power levels (e.g. $> -62\,$dBm for 20\,MHz channels~\cite{gast-2017}) of the received LTE-U signal received at the WiFi AP, the WiFi transmitter will be able to sense the LTE-U signal using its ED-based physical CS mechanism
and therefore defer from the channel during the LTE-U \textit{ON} phase. Hence only during the \textit{OFF} phase the channel is used by WiFi. For lower LTE-U signal power levels (e.g. $< -62\,$dBm for 20\,MHz channels) the ED-based CS mechanism of Wi-Fi is unable to detect any ongoing LTE-U transmission and hence Wi-Fi will also transmit during the LTE-U \textit{ON} phase resulting in potential packet corruption due to too high co-channel interference at the receiver, i.e. inter-technology hidden node problem. In the worst case, any WiFi packet being transmitted during the LTE-U \textit{ON} phase will get lost (also retransmissions) and only packet transmissions during the \textit{ON} phase are successful.

In summary, we can identify three different interference regimes for WiFi:
\begin{enumerate}
	\item strong interference level, i.e. above ED-CS threshold,
	\item medium interference level, i.e. below ED-CS threshold but with noticeable impact,
	\item weak to no interference, i.e. no noticeable impact
\end{enumerate}
These three interference regimes, marked with numbers 1 to 3, are visible in the results presented in Fig.~\ref{fig:udp_degradation}.

%% file: sections/problem.tex
%
%
\section{Problem Statement}

Our main goal is to estimate the effective available medium airtime of each link in a WiFi BSS under LTE-U interference. This can be formulated as follows:

\medskip

\noindent \textbf{\textit{Instance}}: A WiFi BSS with AP $A$ and a set of associated STAs $w \in \mathcal{W}$. A set of $\mathcal{L}$ LTE-U BS nodes with identical and aligned duty cycles. 

\medskip

\noindent \textbf{\textit{Objective}}: The goal is to find the effective available medium airtime at the physical layer for each DL and UL connection of the WiFi BSS, $C_{A,w}$ and $C_{w,A}$, by taking into account external interference from LTE-U BSs. For DL from AP $A$ to STA $w$ we have:
\begin{align}\label{eq:airtime_eq}
C_{A,w} = \frac{1}{\uptau} \sum_{t=0 \ldots \uptau}{X_t^{A,w}}
\end{align}

\noindent where $\uptau$ is the LTE-U period. Note, the UL direction, $C_{w,A}$, is defined analogously. Variable $X_{t}^{A,w}$ takes a value of zero or one depending on whether a WiFi transmission on the link $A \rightarrow w$ on time slot $t$, i.e. OFDM symbol, is blocked/interfered by the \textit{ON} phase of LTE-U BS(s) or not:
\begin{align}\label{eq:assignment_eq}
X_{t}^{A,w} = 
\begin{cases}
    0, & \text{if } \sum_{L \in \mathcal{L}}{\mathrm{Prx}_A^L} \geq -62\,\mathrm{dBm} \\
		0, & \text{if } \tilde{\gamma}^{\mathcal{L}}_{A,w} < \tilde{\gamma}_{A,w} - \psi \vee  \tilde{\gamma}^{\mathcal{L}}_{w,A} < \phi \\
    1, & \text{otherwise}
\end{cases}
\end{align}

\noindent here the first condition considers blocking WiFi transmission due to sensing LTE-U signals, i.e. received power from LTE-U BS(s) $\mathrm{Prx}_A^L$ is larger the ED threshold. Second condition is for corrupted transmission due too strong interference from LTE-U BS(s) resulting in too low SINR for either data, $\tilde{\gamma}^{\mathcal{L}}_{A,w}$, or Acknowledgment (ACK) packet, $\tilde{\gamma}^{\mathcal{L}}_{w,A}$, during the LTE-U \textit{ON} phase. Note, $\tilde{\gamma}_{A,w}$ and $\tilde{\gamma}_{w,A}$ represent the average SNR of the link during the LTE-U \textit{OFF} phase, i.e. no interference from LTE-U. 
The rationale behind this is the observation that rate control algorithms like Minstrel~\cite{minstrel2005} aim to optimize for maximum expected throughput and hence adapt the Modulation and Coding Scheme to the channel quality during the larger \textit{OFF} phase. Here $\psi$ and $\phi$ are the margin and minimum SINR for reception of ACK packets respectively.

%% file: sections/approach.tex
%
%
\section{WiPLUS's Design \& Implementation}

\subsection{Requirements}

We believe that the proposed solution should meet the following requirements:
\begin{itemize}
	\item Real-time estimation of the effective available medium airtime for all DL/UL connections in a WiFi BSS,
	\item Accurate estimation in all three LTE-U interference regimes (cf. Sec.~\ref{sec:under}),
	\item Passive approach without any active measurements, i.e. zero overhead in radio channel,
	\item Low computational complexity \& cheap solution,
	\item Simple software installation process at WiFi AP and no need for modification of WiFi STAs,
	\item Realization using commodity 802.11 hardware and no need for additional hardware
\end{itemize}

\subsection{Approach}

To be a fast and lightweight solution, WiPLUS utilizes a \textbf{three step approach}. The first step is used to detect solely the existence of any interference from LTE-U on a WiFi BSS with the lowest possible complexity. In case LTE-U interference is detected, in a subsequent second step, the effective available medium airtime on each link, DL from AP to STAs, as well as timing information about the position of the LTE-U \textit{ON} and \textit{OFF} phases are estimated. Note that in WiPLUS the UL cannot be directly estimated as WiPLUS is not executed on the STAs; instead it is assessed from the DL. The third step involves interference mitigation strategies such as interference-aware scheduling of packet transmissions and rate and channel bonding adaptation. The three steps are executed directly and in nearly real-time on the WiFi AP. 

\iftoggle{techreport}{
	\begin{wrapfigure}[8]{R}{0.5\textwidth}
		\centering
		\includegraphics[width=0.8\linewidth]{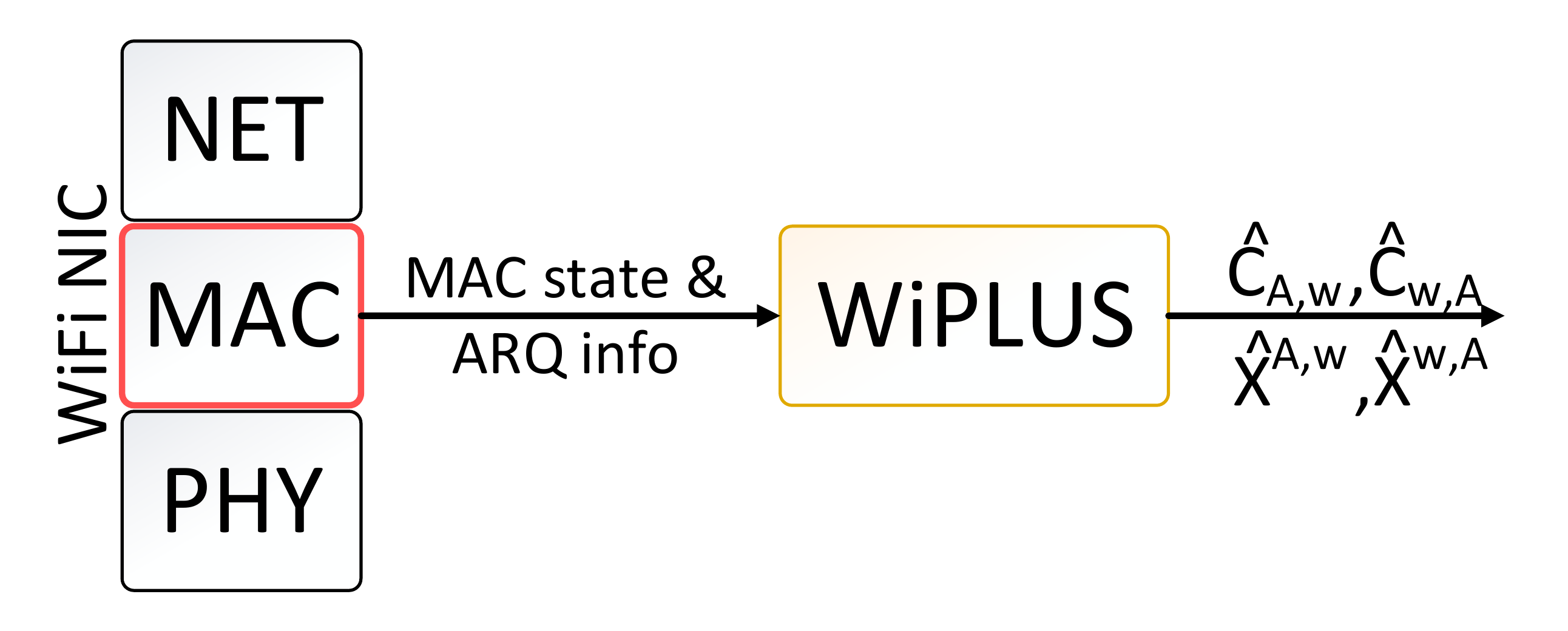}
		\caption{WiPLUS is based on MAC layer monitoring.}
		\label{fig:wiplus_mac_monitoring}
	\end{wrapfigure}	
}%
{
}

WiPLUS obtains all required input data by just performing \textbf{MAC layer monitoring} (Fig.~\ref{fig:wiplus_mac_monitoring}) and hence is therefore fully passive. As the 802.11 MAC is a finite state machine (FSM) and the 802.11 MAC Automatic Repeat reQuest (ARQ) functionality tracks information about frame retransmissions, the sampling of these MAC FSM state transitions and ARQ information is used by WiPLUS to detect an interfering LTE-U signal within all three interference regimes. In a nutshell, this is done as follows. As WiFi cannot decode LTE-U frames it has to rely on ED-based carrier sensing. Hence, the LTE-U’s share of medium time equals the time share that corresponds to energy detection without triggering packet reception (interference regime one, Sec.~\ref{sec:under}). Unfortunately, if the LTE-U signal is in interference regime two (below ED CS threshold), it can, without being detected by Wi-Fi’s ED CS, corrupt ongoing WiFi transmissions. However, as WiPLUS observes the MAC ARQ state, i.e. counting the number of MAC layer retransmissions to detect packet corruption (either the data or the ACK frame), it is able to detect the LTE-U signal even in interference regime two.

\iftoggle{techreport}{
}%
{
	\begin{figure}[h!]
		\centering
		\includegraphics[width=0.5\linewidth]{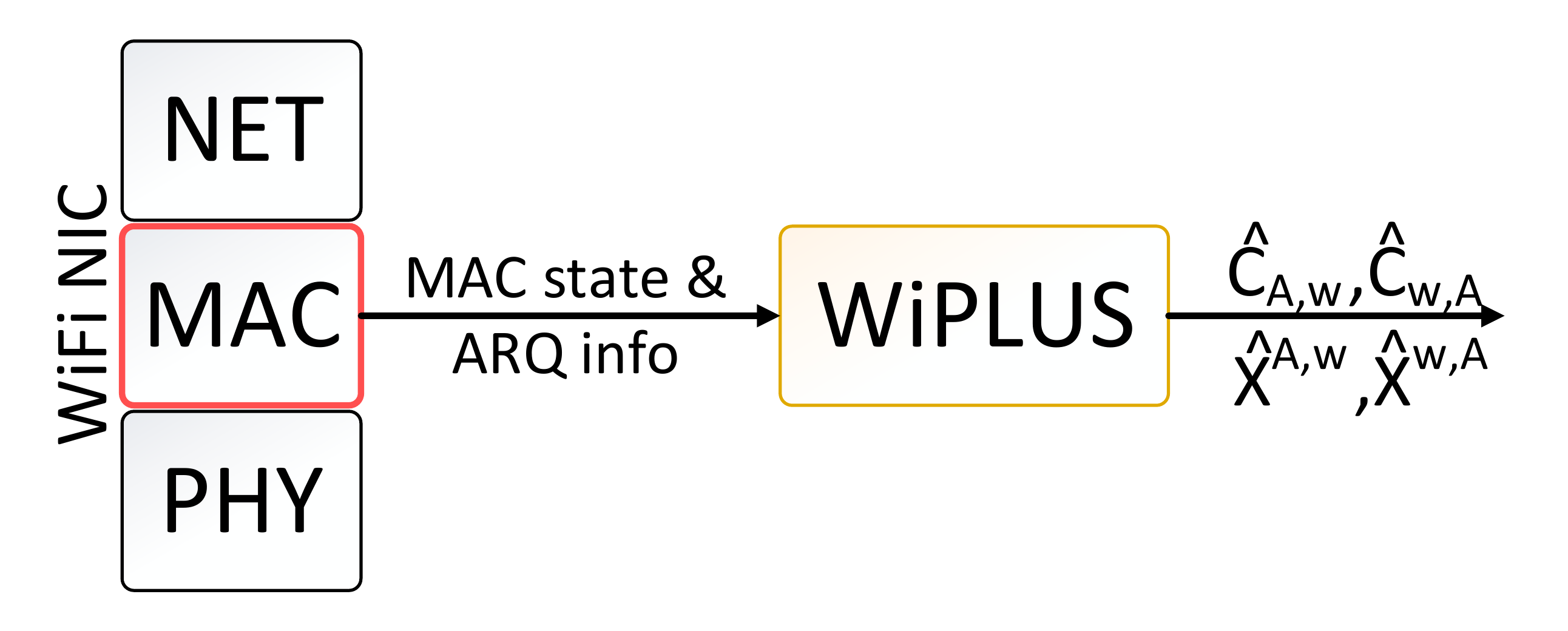}
		\caption{WiPLUS is based on MAC layer monitoring.}
		\label{fig:wiplus_mac_monitoring}
	\end{figure}	
}

\begin{figure*}[t!]
	\centering
	\includegraphics[width=1\linewidth]{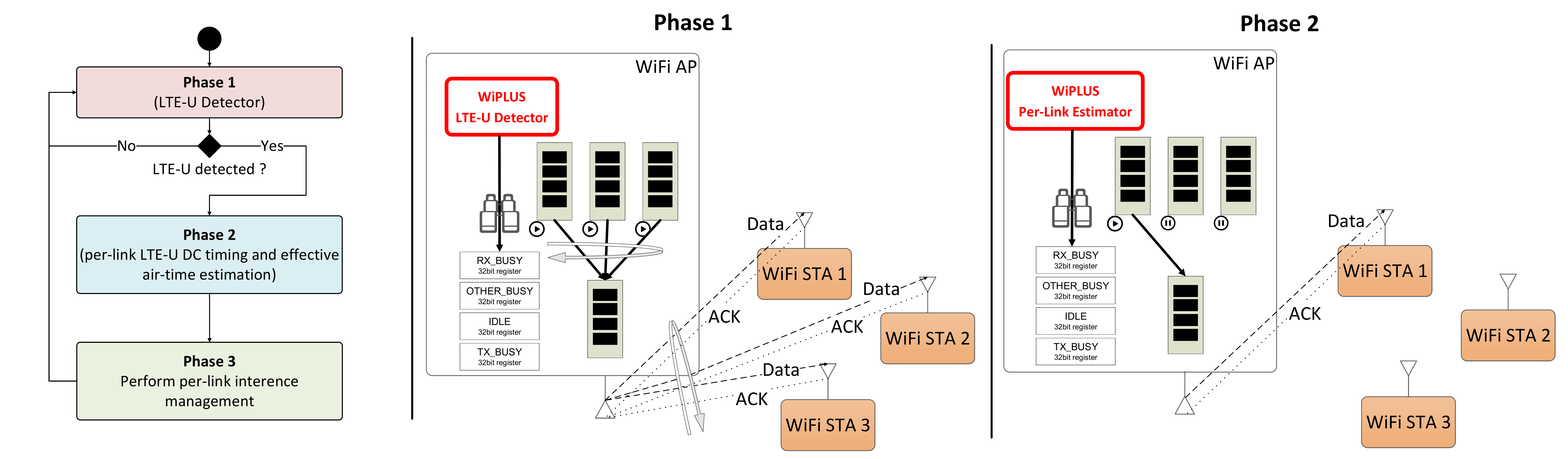}
	\caption{WiPLUS consists of three phases --- Phase 1: LTE-U detector runs passively in the background and terminates in case the presence of any interfering LTE-U signal is detected (middle). Phase 2: In order to be able to discriminate the interference level on each DL connection WiPLUS switches into a time slotted DL access. Here each link can be tested independently so that its effective available medium airtime as well as precise timing information about the LTE-U \textit{ON}/\textit{OFF} phases can be estimated (right). Phase 3: execution of interference mitigation strategies.}
	\label{fig:wiplus-2step}
\end{figure*}

In more detail, WiPLUS requires the following input data describing the state of the MAC FSM and ARQ information: 
\begin{enumerate}
	\item \textit{TX\_BUSY}: total MAC time spent in transmit state,
	\item \textit{RX\_BUSY}: total MAC time spent in receive state,
	\item \textit{OTHER\_BUSY}: total MAC time spent in energy\_detection state, i.e. energy detected without triggering WiFi packet reception (not entering MAC receive state),
	\item \textit{IDLE}: total MAC time in idle state, 
	\item \textit{ACK\_FAIL}: ARQ state, i.e. number of frames being not acknowledged.
\end{enumerate}

WiPLUS's signal detection component samples the MAC layer periodically in order to obtain these data (Fig.~\ref{fig:wiplus_mac_monitoring}). Following this, further signal processing and filtering enables to calculate the duty cycle of the interfering LTE-U signal from which the effective medium airtime for WiFi is derived, cf. Sec.\ref{sec:detailed}.


By observing the MAC state machine it is only possible to detect the presence of an interfering LTE-U signal, but it is not possible to distinguish which link is being impaired by the LTE-U interference. To overcome this limitation, WiPLUS uses this approach only for the detection of an interfering LTE-U signal and starts, after an interfering LTE-U signal was detected, 
to measure each downlink connection separately, e.g. by separating all links in time.\\
In this slotted mode, all links are not served round robin as it is done usually, but rather the sending of frames of the single links is done timely separated, e.g. in slot 1 link A is served only, and in slot 2 link B is served only. 

Using this approach, it is possible to map the observations from the ARQ and MAC FSM to distinct slots, which therefore allows to monitor every link separately and to discriminate the impact of LTE-U interference on each link (Fig.~\ref{fig:wiplus-2step}).
%
%
%
\subsection{Algorithm Design and Signal Processing}\label{sec:detailed}

Fig.~\ref{fig:lte-u-flow-chart} shows a flow chart of WiPLUS describing the steps involved in estimating the LTE-U duty-cycle from which the effective available medium airtime for WiFi is derived:
\iftoggle{techreport}{
	\begin{wrapfigure}[17]{R}{0.5\textwidth}
		\centering
		\includegraphics[width=\linewidth]{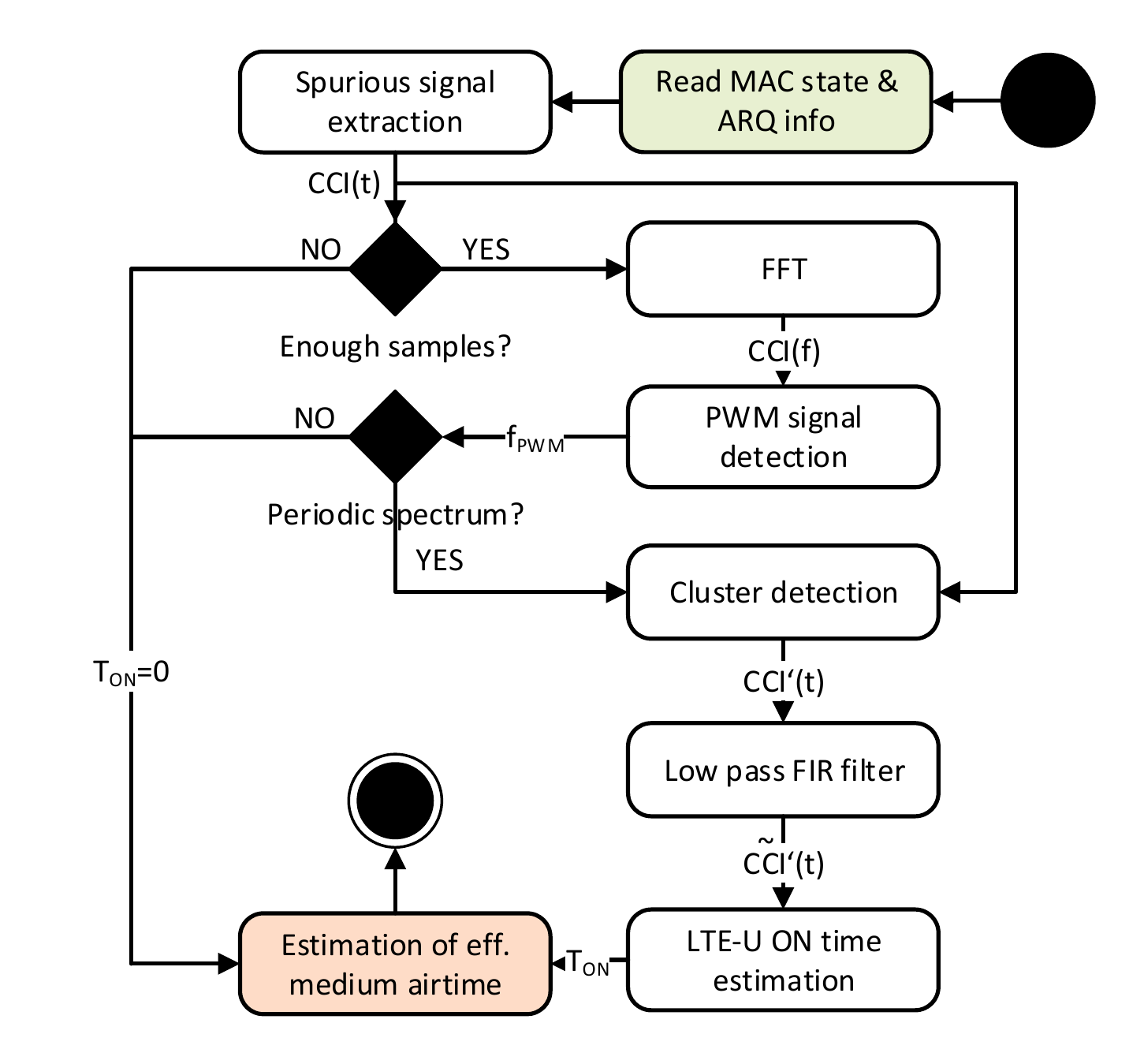}
		\caption{Flow chart of LTE-U duty cycle detector.}
		\label{fig:lte-u-flow-chart}
	\end{wrapfigure}	
}%
{
	\begin{figure}[h!]
		\centering
		\includegraphics[width=0.7\linewidth]{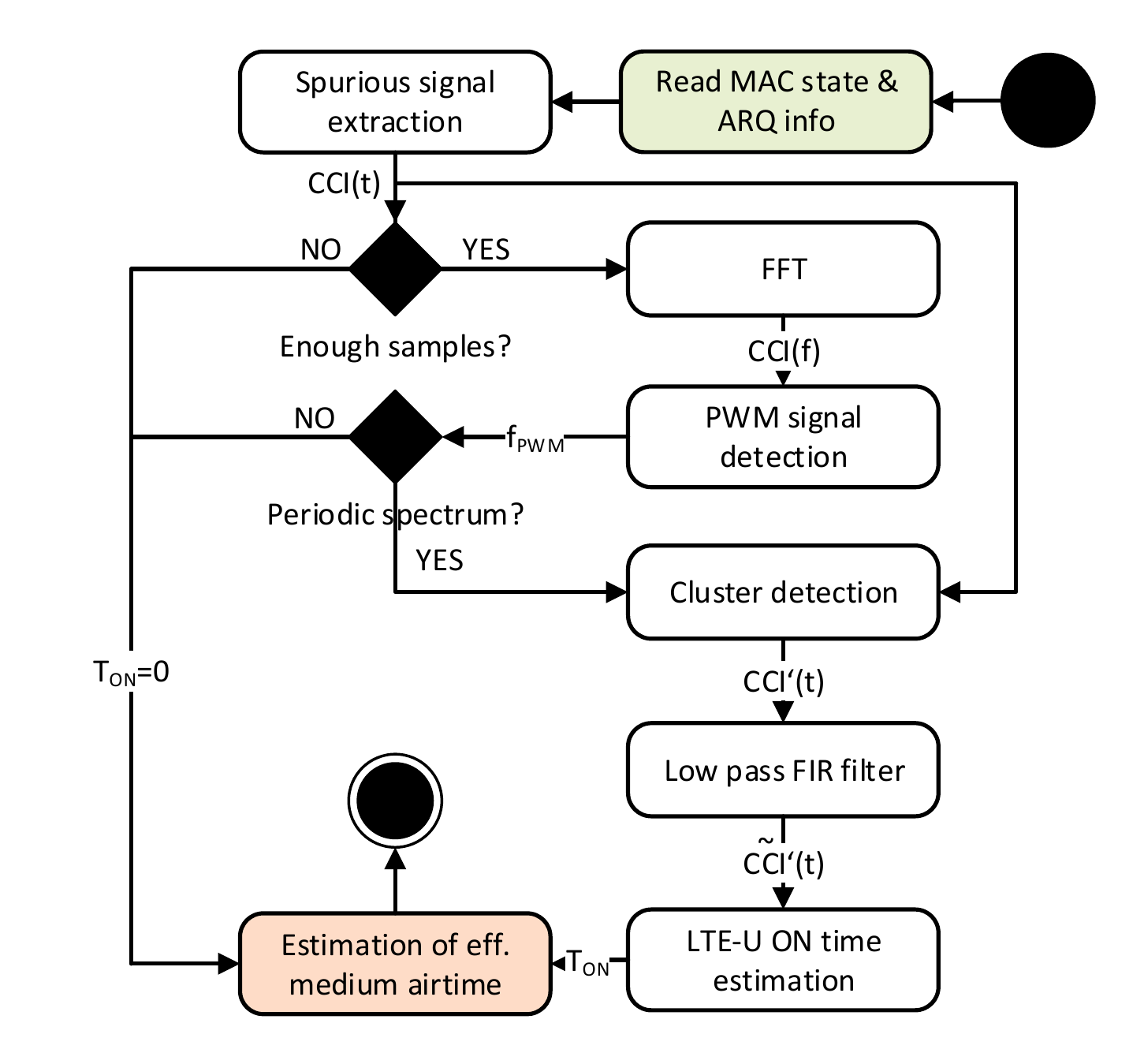}
		\caption{Flow chart of LTE-U duty cycle detector.}
		\label{fig:lte-u-flow-chart}
	\end{figure}	
}

\textbf{Step 1 -- raw data acquisition:} The input for the first step is data from the 
MAC FSM states and the ARQ functionality.
\begin{align}\label{eq:raw_data_input}
S_t^{\mathrm{TX}}, S_t^{\mathrm{RX}}, S_t^{\mathrm{OTHER}}, S_t^{\mathrm{ACK\_FAIL}}&, \forall t \in 0 \ldots W
\end{align}
where $S_t^{\mathrm{TX}}$, $S_t^{\mathrm{RX}}$ and $S_t^{\mathrm{OTHER}}$ are the relative dwell times for \textit{TX\_BUSY}, \textit{RX\_BUSY} and \textit{OTHER\_BUSY} (e.g. $S_t^{\mathrm{RX}} = 50$ if the MAC stayed 50 percent of the corresponding sample $t$ in the RX state), while $S_t^{\mathrm{ACK\_FAIL}}$ is the number of failed transmissions during sample $t$.

\textbf{Step 2 -- spurious signal extraction:} In the next step we apply signal extraction in which we aim for filtering out noise. For this reason we create a new signal $R_t$:
\begin{align}\label{eq:signal_extraction}
R_t = 
\begin{cases}
S_t^{\mathrm{OTHER}}, & \text{if } S_t^{\mathrm{TX}}=S_t^{\mathrm{RX}}=0, \\
S_t^{\mathrm{TX}}, & \text{if } S_t^{\mathrm{ACK\_FAIL}}>0, \\
S_t^{\mathrm{TX}}, & \forall t': t' \geq t: S_{t'}^{\mathrm{ACK\_FAIL}}>0 \ \wedge \\
& S_{t \ldots t'}^{\mathrm{TX}} = 100\\
0, & \text{otherwise}
\end{cases}
\end{align}
While the first case filters out noisy samples, the second and third account for interfered transmission attempts, i.e. unsuccessful transmission due to corruption of either data packet or acknowledgement frame, spanned over single and multiple samples respectively. Note, in the third condition we look-ahead: $R_t$ takes the value of $S_{t}^{\mathrm{TX}}$ if we observe a series of samples with $S_{t}^{\mathrm{TX}} = 100$ followed by an ACK fail.

Moreover, we abort the process in case not enough samples could be extracted, $<1\,$\% of samples, i.e. $\frac{\sum_{t \in 1 \ldots W}{\text{sign}(R_t)}}{W} \leq 0.01$.
This threshold was found to be feasible during our experiments and
increases the robustness of the algorithm, thus reducing the probability of false detections.

\textbf{Step 3 -- FFT / PWM signal detection:} In the next step we compute the normalized power spectrum of the signal $R_t$ using the fast Fourier transform (FFT) algorithm. This is used
as input for a peak detector to find the fundamental frequency, $f_{\mathrm{PWM}_0}$, and second/third harmonics, $f_{\mathrm{PWM}_1}$ and $f_{\mathrm{PWM}_2}$. We abort in case no periodic spectrum can be detected.

\textbf{Step 4 -- clustering:} We use KMeans clustering on $R_t$ to detect cluster centers in time domain. Therefore, we performed clustering for different values for $k = \left\lceil \frac{1}{f_{\mathrm{PWM}_0}} \right\rceil-2 \ldots \left\lceil \frac{1}{f_{\mathrm{PWM}_0}} \right\rceil+2$. We determined the optimal K by silhouette analysis around desired K.
For all $N$ points in a cluster the distance $d_{i,k}$ to the cluster center is calculated, such that $D_{k} = \{d_{i,k} | i=1...N\}$.
We assume that all points in a cluster are 
uniformly distributed in time.
Finally, we set all signal parts outside $2 \times \text{median}(D_{k})$ around cluster centers to zero in order to suppress outliers.

\textbf{Step 5 -- low pass filtering:} We low pass filter the remaining signal to overcome possible imperfections of the KMeans algorithm and to bridge possible gaps, i.e. smoothing.
Therefore, $f_{\mathrm{PWM}_0}$ is used as filter cutoff frequency. We estimate the 
effective LTE-U ON time, $T_{\mathrm{ON}}$, by computing the mean segment duration of the remaining parts in $R_t > 0$ (suppress outliers).

\textbf{Step 5 -- calculate effective available medium air-time:} Finally, we are able to compute the effective available medium air-time which is: $\hat{C} = 1 - T_{\mathrm{ON}} \times f_{\mathrm{PWM}_0}$.


%
%
%
\subsection{Prototype Implementation Details}

For our prototype we selected WiFi chipsets based on Atheros AR95xx as they allow monitoring of the MAC state machine and provide information about the ARQ state. Fig.~\ref{fig:regmon} illustrates the structure of the signal detection logic of Atheros based WiFi chips. The three most interesting building parts such as weak signal detection, strong signal detection and energy detection are depicted. For strong signal detection, it is determined that a signal may exist by the arrival of a stronger signal necessitating a drop in receive gain (“capture effect”). For weak signal detection, it is determined that a signal may exist due to a sudden increase in measured in-band power, followed shortly by a correlation-based algorithm that uses the structure of the preamble signal. Both detectors are triggers to switch the MAC state machine into receive state, while the pure ED without strong or weak signal detection puts the MAC into the other busy mode \cite{patent-atheros1}.

\iftoggle{techreport}{
	\begin{wrapfigure}[15]{L}{0.5\textwidth}
		\centering
		\includegraphics[width=1\linewidth]{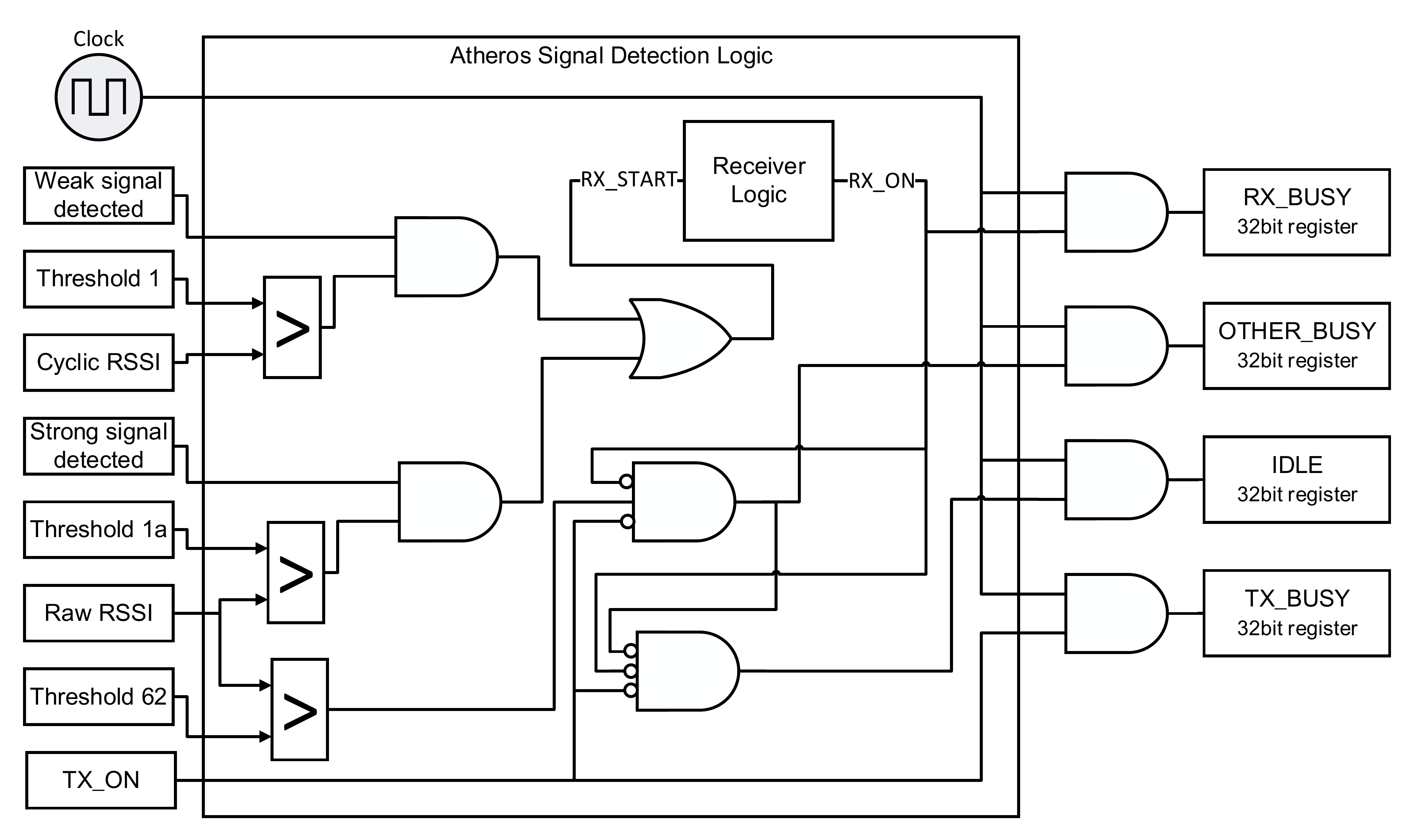}
		\caption{Block diagram of Atheros signal detection logic and MAC state registers (adapted from~\cite{patent-atheros1}).}
		\label{fig:regmon}
	\end{wrapfigure}	
}%
{
	\begin{figure}[h!]
		\centering
		\includegraphics[width=0.89\linewidth]{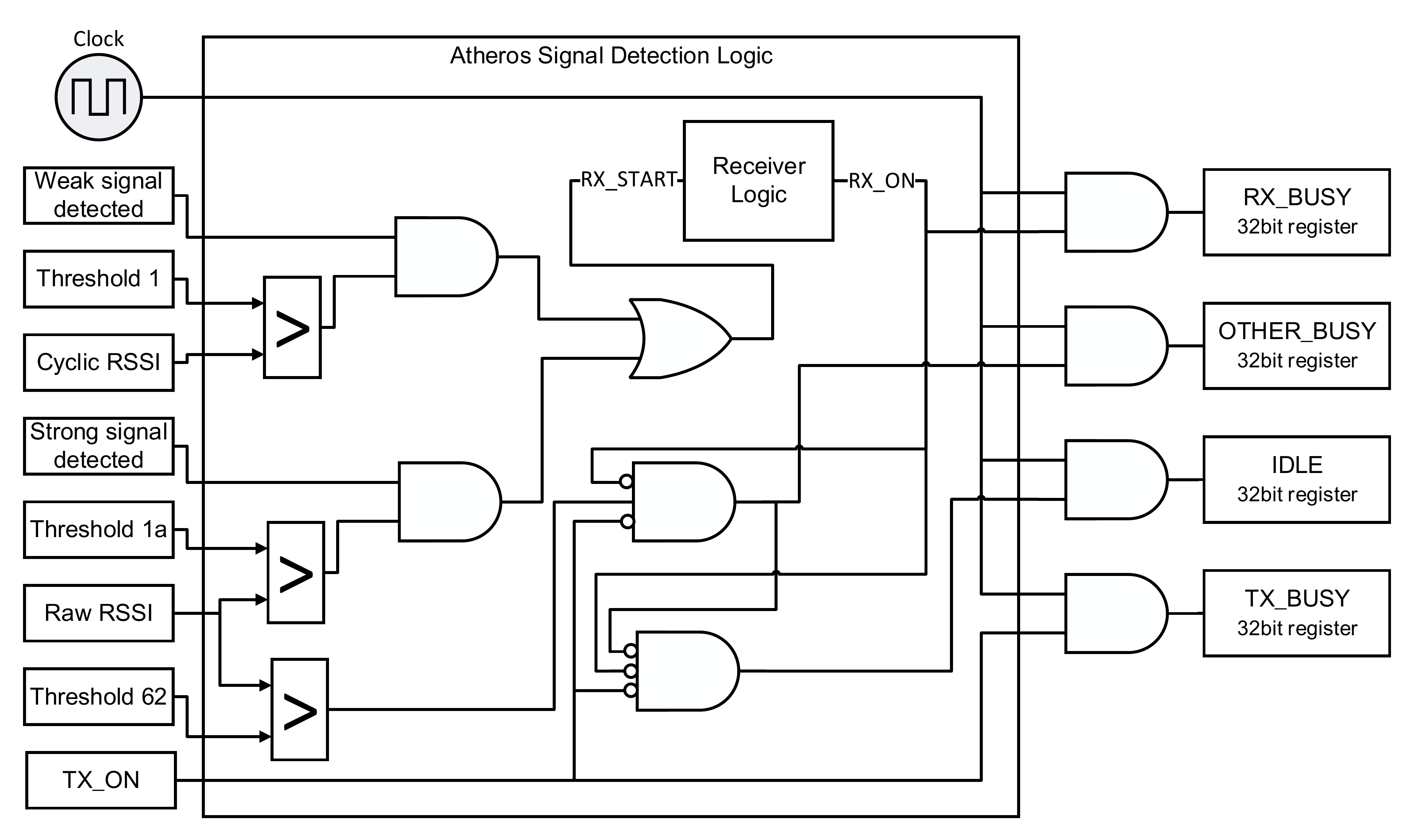}
		\caption{Block diagram of Atheros signal detection logic and MAC state registers (adapted from~\cite{patent-atheros1}).}
		\label{fig:regmon}
	\end{figure}	
}

Moreover, on Atheros platforms, all MAC states are connected to 32\,bit registers, which sample the current state using a 40\,MHz clock, cf. Fig.~\ref{fig:regmon}. 

During the first two phases of WiPLUS, we sample the MAC state registers with a rate of 2\,kHz and process the data in chunks of 1\,s window sizes, i.e. $W=2000$ samples. For getting access to the registers of the ATH9K MAC FSM and ARQ functionality, we use a modified version of the regmon tool~\cite{Huehn2013}. For the actual signal processing we rely on Python libraries (NumPy, SciPy). Further, to enable queue-control, cf. Fig.~\ref{fig:wiplus-2step}, we use a modified version of the ATH9k hMAC~\cite{zehl16hmac} tool, which allows us to pause and un-pause the software queues of the ATH9K driver originally used for power saving and frame aggregation.

%% file: sections/apps.tex
%
%

\section{WiPLUS's Interference Management Applications}

\iftoggle{techreport}{
}%
{
\begin{figure*}[t!]
\centering
\includegraphics[width=1\linewidth]{figs/wiplus_apps_em}
\caption{WiPLUS apps: (1) Interference-aware Channel Assignment, (2) Interference-aware Load Balancing Clients, (3) Interference-aware Medium Access, (4) Interference-aware Client Association Steering, (5) Interference-aware Channel Bonding, (6) Interference-aware Rate Adaptation.}
\label{fig:wiplus_apps}
\end{figure*}
}

In its third step WiPLUS performs interference management in the WiFi network. It is executed after an interfering LTE-U signal was detected and the affected links have been identified. This section gives an overview about possible interference mitigation strategies.


\subsection{Interference-aware Channel Assignment}

\iftoggle{techreport}{
\begin{figure*}[t!]
\centering
\includegraphics[width=1\linewidth]{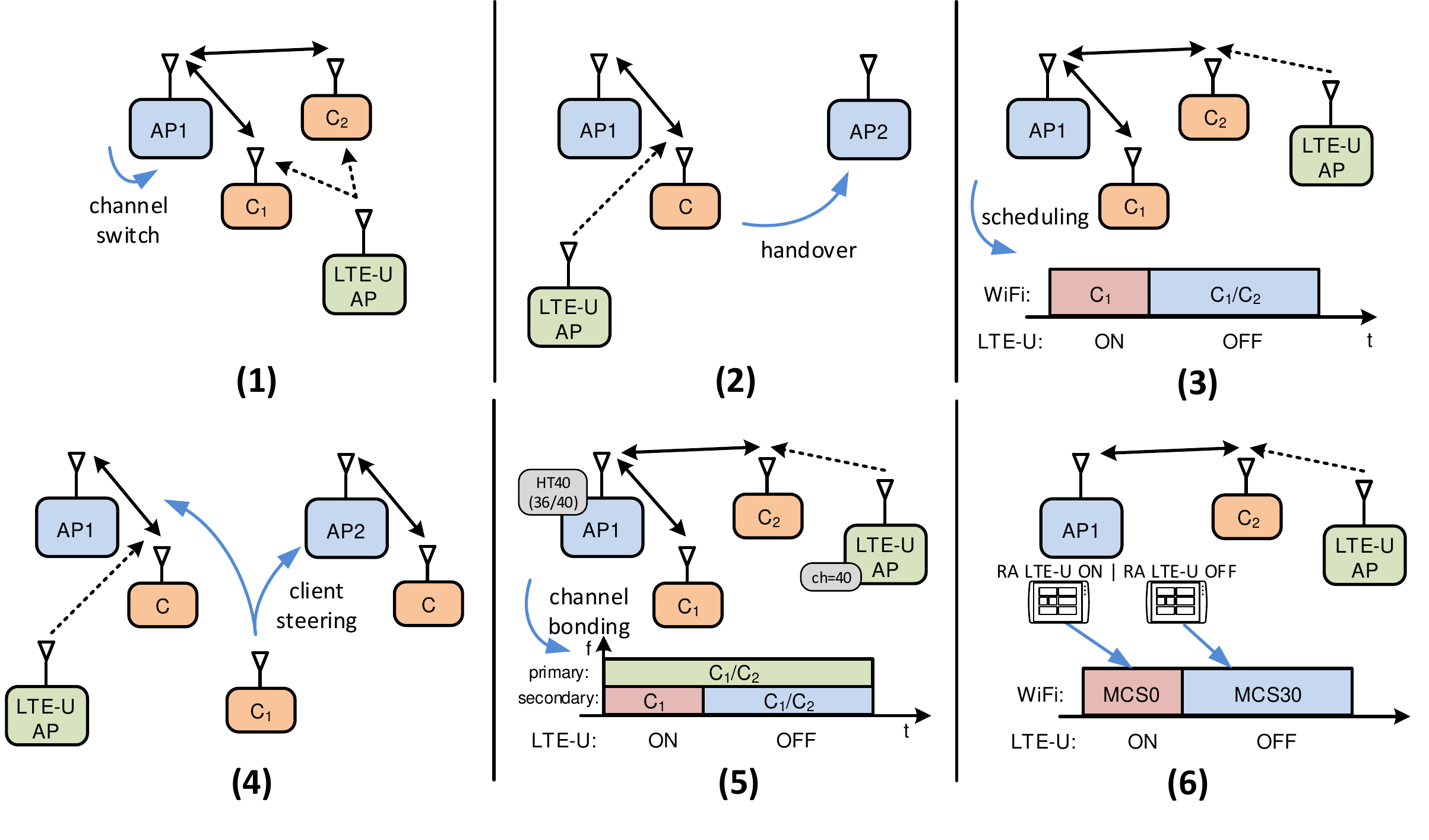}
\caption{WiPLUS apps: (1) Interference-aware Channel Assignment, (2) Interference-aware Load Balancing Clients, (3) Interference-aware Medium Access, (4) Interference-aware Client Association Steering, (5) Interference-aware Channel Bonding, (6) Interference-aware Rate Adaptation.}
\label{fig:wiplus_apps}
\end{figure*}
}%
{
}

The simplest approach for WiFi to mitigate interference from LTE-U is to abandon the affected channel (Fig.~\ref{fig:wiplus_apps}.1). It is especially meaningful if the LTE-U interference degrades the performance of the majority of links in a WiFi BSS.
Such a technique can be applied in Enterprise networks and in home networks with additional usage of e.g.~\cite{resfi}.
By exploiting the 802.11 DFS functionality such a channel switch can be performed seamless, e.g. as demonstrated in~\cite{zehl2016bigap}.

\subsection{Interference-aware Load Balancing of Clients}

In general radio channels are assigned to WiFi APs on a medium to long-term basis. However, in order to quickly adapt to variations in external interference, i.e. changes in LTE-U duty cycle or network load, the STA associations across APs can be optimized by performing seamless handover operations~\cite{Zubow16bigap_seamless_handover} (Fig.~\ref{fig:wiplus_apps}.2). Such a technique can be applied in Enterprise networks. For home networks this might be possible if the deployment supports cooperation between neighbor APs, e.g. using ResFi~\cite{resfi} together with NxWLAN~\cite{gawlowicz2016nxwlan}.

\subsection{Interference-aware Medium Access}\label{sec:hmac}

Spectrum is a rare resource and it is therefore foreseeable that in future, the 5\,GHz band will become as crowded as the 2.4\,GHz band is now. Therefore, an interference-aware scheduling of packet transmissions might be applied where the WiFi channel access is limited to the LTE-U \textit{OFF} period (Fig.~\ref{fig:wiplus_apps}.3). Note such a scheme is feasible as WiPLUS provides timing information about the detected LTE-U \textit{ON} and \textit{OFF} phases enabling to synchronize the medium access of WiFi with the LTE-U cycle. From practical point of view this can be achieved using the slotted (TDMA) based channel access proposed in~\cite{zehl16hmac}.

\subsection{Interference-aware Client Association Steering}

Another option is to directly influence the STA association process in WiFi. Therefore, each WiPLUS enabled AP can independently estimate the amount the channel is occupied by external interference and announce that value in its beacon frames, e.g. using beacon stuffing as suggested in \cite{zehl2016lows}. An 802.11u/k compliant STA would utilize that value in the AP selection process (Fig.~\ref{fig:wiplus_apps}.4).

\subsection{Interference-aware Channel Bonding}

Channel bonding allows 802.11 devices to operate with channel widths of 40, 80 or even 160\,MHz. Therefore, the primary channel is aggregated with multiple secondary channels. The following interference mitigation scheme can be applied in case external interference is detected on one of the secondary channels. The timing information provided by WiPLUS allows synchronizing the medium access of WiFi with the LTE-U cycle so that during the LTE-U \textit{ON} phase the channel bonding can be restricted to non-interfered secondary channels (Fig.~\ref{fig:wiplus_apps}.5). This can be achieved e.g. using~\cite{Nishat-2016} which allows assigning different channel widths on a per-frame basis.

\subsection{Interference-aware Rate Adaptation}

Rate adaption algorithms like e.g. Minstrel~\cite{minstrel}, are adapting the bitrate (MCS) used by the sender to match the wireless channel conditions, to achieve e.g. best possible throughput. As current rate adaption algorithms are not designed to handle periodic interference, the usage of two rate adaption algorithms might be promising, i.e. one for LTE-U \textit{OFF} and another for the \textit{ON} phase (Fig.~\ref{fig:wiplus_apps}.6).

%% file: sections/eval.tex
%
%
\section{Experiments}\label{sec:exp_sec}

WiPLUS was prototypically implemented and evaluated by means of experiments.

\subsection{Methodology}

We set-up a single WiFi AP with associated STA and a co-located LTE-U BS. The distance between each pair of nodes was set to 3\,m, i.e. triangle, so that the signal from the LTE-U BS was received at same power level by both AP and STA. For WiFi we used Atheros AR95xx network cards and a patched ATH9k driver\footnote{https://github.com/szehl/ath9k-hmac, https://github.com/thuehn/RegMon}. The LTE-U BS waveform was precomputed offline using Matlab and afterwards generated using R\&S SMBV100A Vector Signal Generator. 

The WiFi mode was set to 802.11a and channel 48 (5240\,MHz, U-NII-1) was used. Furthermore, we used only a single antenna for WiFi (no antenna diversity, MIMO, etc.) and also disabled Atheros Adaptive Noise Immunity. The TX power for WiFi was set fixed to 15\,dBm for both AP and STA whereas for LTE-U BS it was varied from +15 to -33\,dBm. For LTE-U two different CSAT cycle lengths, i.e. 80 and 160\,ms, were used. Moreover, the LTE-U duty cycle was set to 33\,\% and a 1\,ms subframe puncturing was used. Moreover, the LTE-U signal was using the same 20\,MHz channel like WiFi.

We set-up a saturated UDP packet flow in the DL from AP to STA. In addition, during the experiment the transmit power of the LTE-U BS signal was varied, i.e. emulating different distances between WiFi BSS and LTE-U BS.

As performance metric we identified the root-mean-square error (RMSE) between the predicted, $\hat{C}_{A,w}$, and the actually available, $C_{A,w}$, medium airtime on the downlink from the WiFi AP $A$ to the STA $w$. The latter was obtained by normalizing the measured actual WiFi UDP throughput with the maximum UDP throughput, i.e. the throughput in absence of LTE-U signal. For both measurements the properties of the RF channel and the experiment nodes were kept constant.

%
%
\subsection{Results}

\noindent \textbf{Experiment 1: (Backlogged LTE-U traffic)} For the LTE-U BS a CSAT cycle length of 80\,ms and a full buffer traffic model is used.

\medskip

\iftoggle{techreport}{
	\begin{wrapfigure}[15]{L}{0.5\textwidth}
		\centering
		\includegraphics[width=1\linewidth]{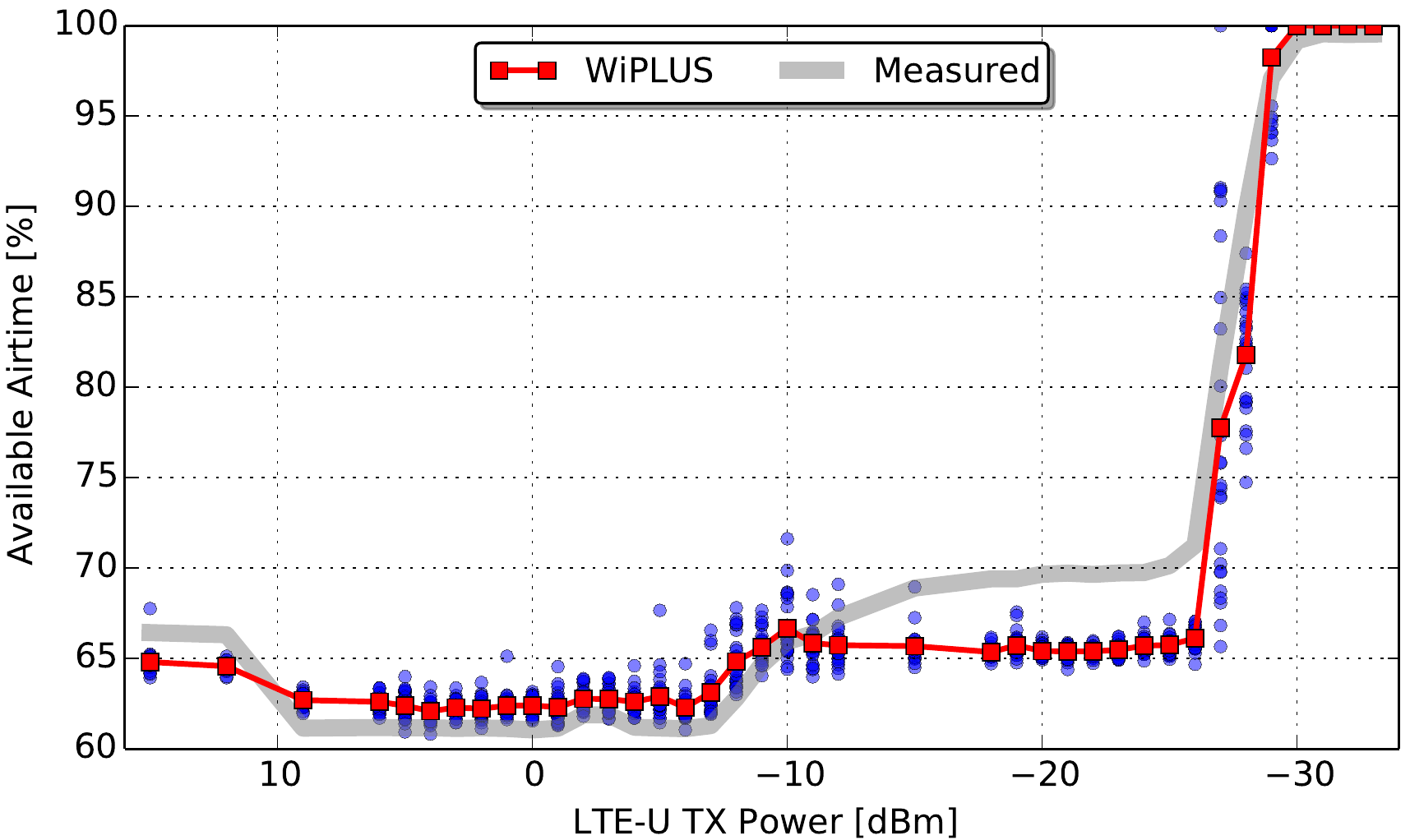} 
		\caption{LTE-U BS with CSAT cycle length of 80\,ms and backlogged traffic.}
		\label{fig:lteu_eval_33}
		
	\end{wrapfigure}	
}%
{
}

\noindent \textbf{Result 1:} The results are shown in Fig.~\ref{fig:lteu_eval_33}. We can clearly see that WiPLUS is able to accurately estimate the effective available medium airtime of the WiFi link even at very low LTE-U TX power levels, i.e. -29\,dBm. For high-power interference, i.e. above ED-CS, the WiFi AP is able to sense the LTE-U signal and the WiFi throughput is reduced (relative to its LTE-U free throughput) by 33\% which corresponds to effective time-sharing. At lower LTE-U interference levels there is a reduction in available medium airtime due to corruption of WiFi packets which again is precisely estimated by WiPLUS. The overall RMSE is around 2.7 percentage points.

\iftoggle{techreport}{
}%
{
	\begin{figure}[h!]
		\centering
		\includegraphics[width=0.95\linewidth]{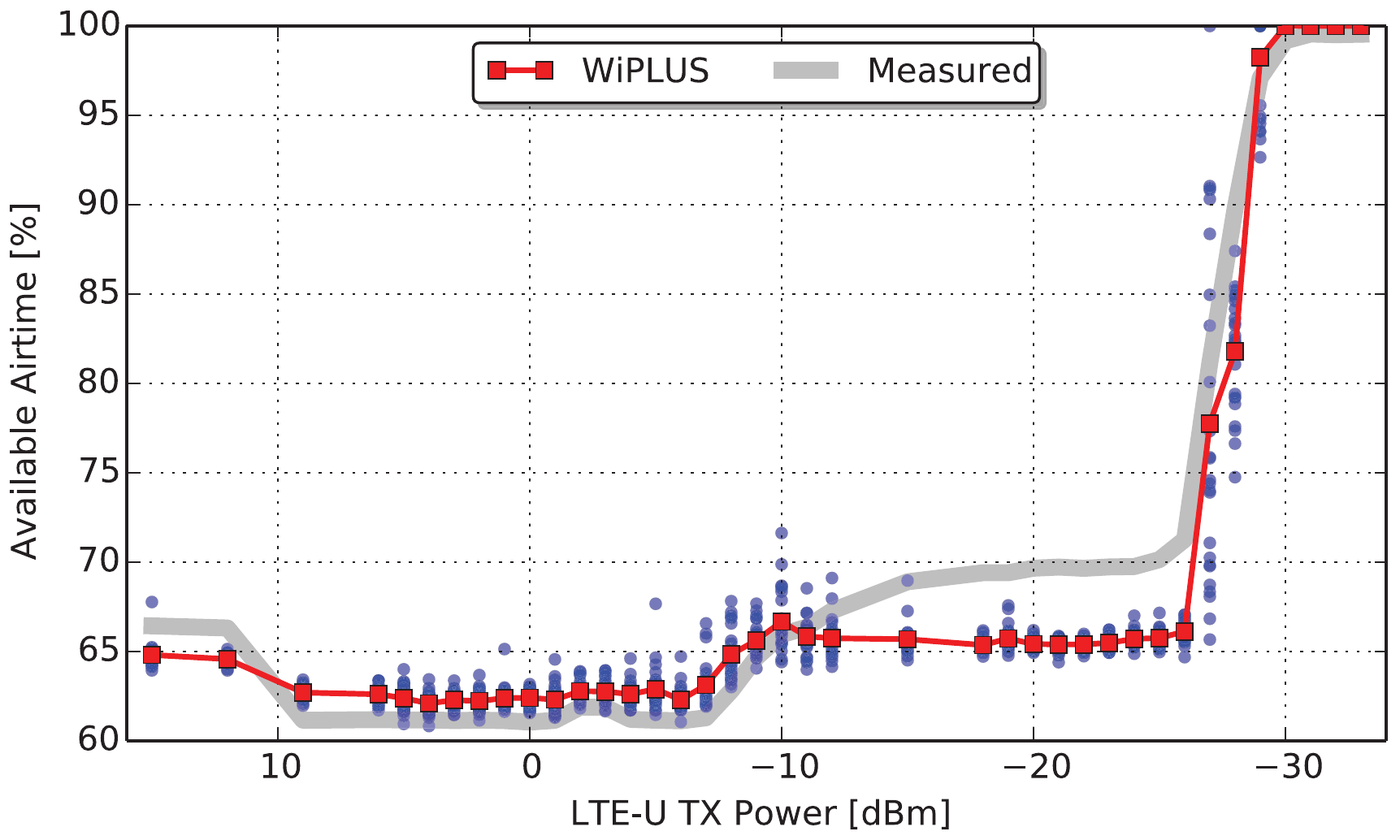} 
		\caption{LTE-U BS with CSAT cycle length of 80\,ms and backlogged traffic.}
		\label{fig:lteu_eval_33}
	\end{figure}
}

\medskip

\noindent \textbf{Experiment 2: (Variable LTE-U traffic)} The setup is as in previous experiment with a fixed duty cycle of 33\% but with variable buffer traffic model, i.e. the LTE-U \textit{ON} phase was loaded uniform random between 30\,\% and 100\,\% which corresponds to an effective LTE-U duty cycle of 16\%.

\medskip

\iftoggle{techreport}{
	\begin{wrapfigure}[14]{R}{0.5\textwidth}
		\centering
		\includegraphics[width=1\linewidth]{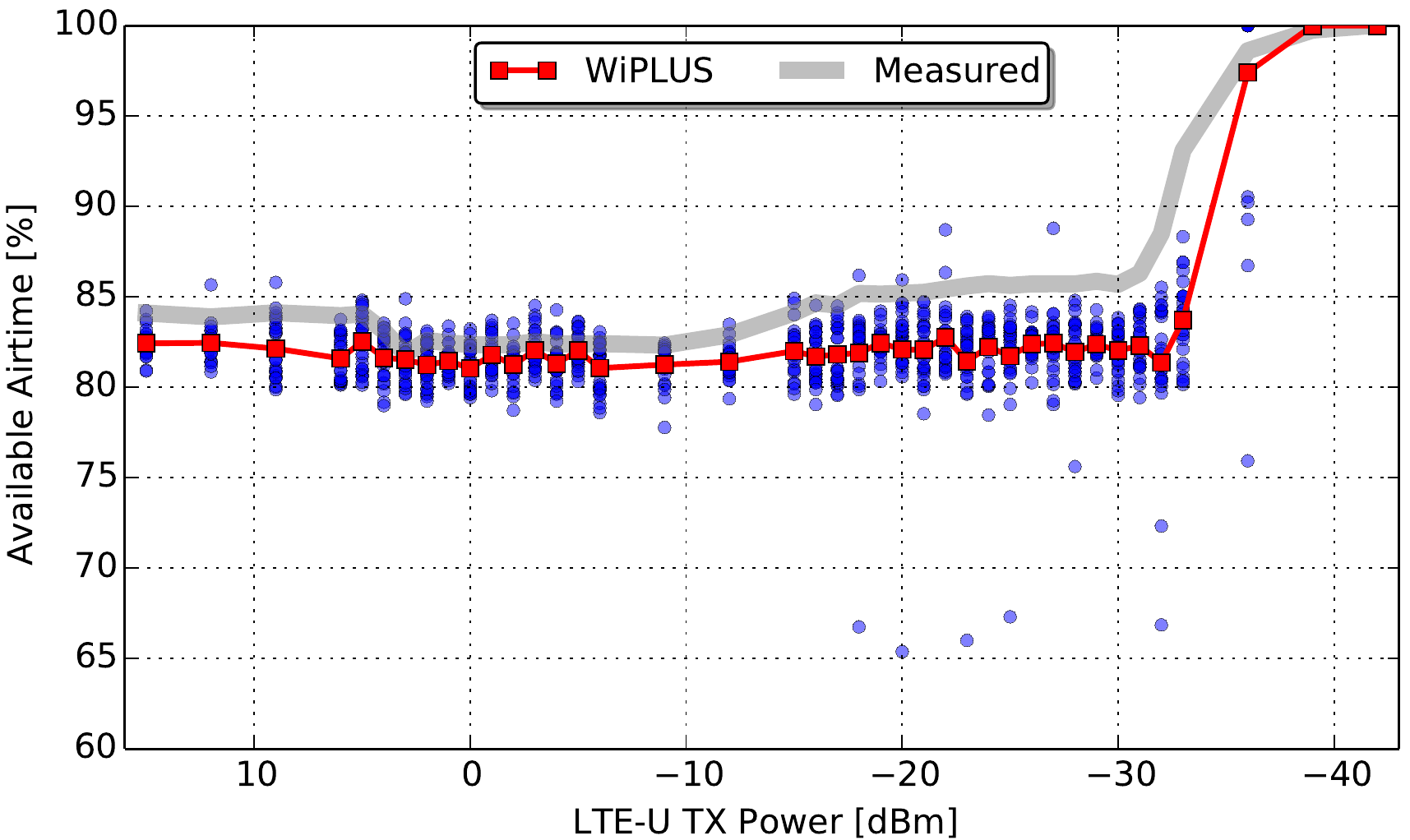} 
		\caption{WiPLUS results for LTE-U BS with CSAT cycle length of 160\,ms.}
		\label{fig:lteu_eval_33_random}
		
	\end{wrapfigure}	
}%
{	
}

\noindent \textbf{Result 2:} The results are shown in Fig.~\ref{fig:lteu_eval_33_random}. We see that WiPLUS is able to accurately estimate the available medium airtime of the WiFi link which corresponds to the effective LTE-U duty cycle.

\medskip
\noindent \textit{\textbf{Takeaways: }}WiPLUS is able to accurately estimate the effective available medium airtime of the WiFi downlink even at very low LTE-U power levels for both backlogged and variable LTE-U traffic.

\iftoggle{techreport}{	
}%
{
	\begin{figure}[h!]
		\centering
		\includegraphics[width=0.95\linewidth]{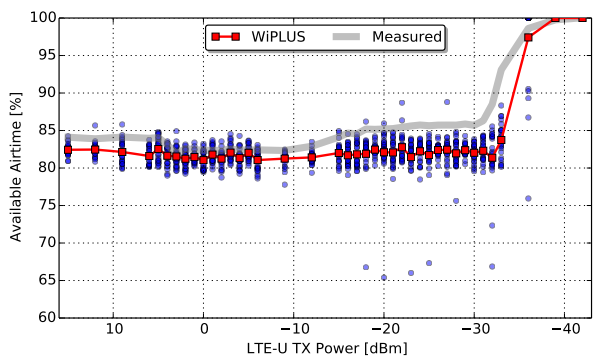} 
		\caption{WiPLUS results for LTE-U BS with CSAT cycle length of 160\,ms.}
		\label{fig:lteu_eval_33_random}
	\end{figure}
	
}

%% file: sections/simulation.tex
%
%
\section{Simulations}

As WiPLUS is running solely on the WiFI AP it has to derive the effective available medium airtime of the UL from the DL measurements resulting in an error between the real $C_{A,w}$ and predicted $\hat{C}_{A,w}$ value. The following two situations can lead to such a misprediction:
\begin{itemize}
	\item The LTE-U BS signal is weak enough so that it cannot be detected by WiPLUS running on the AP side, i.e. below ED-CS and weak enough to not corrupt DL transmission (data plus ACK). Note ACK frames are more robust to interference as they send on the most robust MCS. However, the interference from LTE-U is strong enough to either block the STA from UL channel access (i.e. above ED-CS) or to corrupt the uplink data transmission, i.e. either data or ACK frame. 
	\item The LTE-U signal is strong enough so that it can be detected by STA (i.e. above ED-CA); however, it is not harming the WiFi transmission, i.e. exposed terminal problem.  
\end{itemize}

In the former case WiPLUS \textbf{underestimates} whereas in the latter it \textbf{overestimates} the actual effective available medium airtime.

\subsection{Methodology}

We performed system-level simulations using Matlab according to the methodology recommended by the IEEE 802.16m group~\cite{zhuang09emd}. 

\iftoggle{techreport}{
	\begin{wraptable}[13]{L}{8cm}
		\footnotesize
		\centering
		\begin{tabular}{p{.45\linewidth}p{.4\linewidth}}
			\colheadbegin \textbf{Parameter} & \textbf{Value} \colheadend
			Center frequency, system BW& 5180\,MHz, 20\,MHz\\
			LTE-U tx power, antenna gain& 24\,dBm, 5\,dBi\\
			WiFi AP TX power, antenna gain& 24\,dBm, 5\,dBi\\
			WiFi STA TX power& 18\,dBm\\
			WiFi (AP/STA) noise figure& 9\,dB\\
			ED CS threshold (LTE-U, WiFi)& -62\,dBm\\
			WiFi PHY/MAC& IEEE 802.11a\\
			max. LTE-U duty cycle& 50\%\\
			LTE-U MAC& CSAT\\
		\end{tabular}
		\caption{Simulation parameters.}
		\label{table:simparams}
		\vspace{-10pt}
	\end{wraptable}	
}%
{
}

The placement is shown in Fig.~\ref{fig:system_model}. The WiFi AP is always placed in the middle whereas the WiFi STAs are placed uniform random with a minimum/maximum distance of $3\,m$ and $50\,m$ from WiFi AP respectively. Moreover, the LTE-U BS is placed uniform random in the bounding box with side length of $120\,m$ centered around the WiFi AP. As path loss model we selected the IEEE 802.16m indoor small office (A1) scenario, adapted to 5\,GHz~\footnote{\url{http://www.ieee802.org/16/tgm/contrib/C80216m-07_086.pdf}}, which describes a random mix of line-of-sight (LOS) and non LOS (NLOS) scenarios. Furthermore, we calculated 12\,dB losses for wall penetration. For NLOS and LOS different Shadowing $\sigma$ were taken: 3.1\,db and 3.5\,db respectively. We explicitly calculated the SNIR taking into account the co-channel interference from the LTE-U BS. The remaining parameters are summarized in Table~\ref{table:simparams}. 

For the simulations we assume that WiPLUS is able to perfectly estimate the effective available medium airtime in the downlink, i.e. $\hat{C}_{A,w} = C_{A,w}$.

Finally, as performance metric we computed for each uplink connection the RMSE between the predicted, $\hat{C}_{w,A}$, and the actually available, $C_{w,A}$, medium airtime.

\iftoggle{techreport}{
}%
{
	\begin{table}[t]
		\footnotesize
		\centering
		\begin{tabular}{p{.45\linewidth}p{.5\linewidth}}
			\colheadbegin \textbf{Parameter} & \textbf{Value} \colheadend
			Center frequency, system BW& 5180\,MHz, 20\,MHz\\
			LTE-U tx power, antenna gain& 24\,dBm, 5\,dBi\\
			WiFi AP TX power, antenna gain& 24\,dBm, 5\,dBi\\
			WiFi STA TX power& 18\,dBm\\
			WiFi (AP/STA) noise figure& 9\,dB\\
			ED CS threshold (LTE-U, WiFi)& -62\,dBm\\
			WiFi PHY/MAC& IEEE 802.11a\\
			max. LTE-U duty cycle& 50\%\\
			LTE-U MAC& CSAT\\
		\end{tabular}
		\caption{Simulation parameters.}
		\label{table:simparams}
		\vspace{-10pt}
	\end{table}
}

\subsection{Results}

The results are depicted in Fig.~\ref{fig:wiplus_ul_link_rmse_1_4}. On the left side we see the RMSE of up to 17 percentage points in case both the WiFi AP and the STAs have the same TX power and antenna gain. On the right side we see the results with TX power and antenna gain configuration as suggested by LTE-U forum (cf. Table~\ref{table:simparams}). Here the RMSE drops below 10. A further reduction can be achieved when solving the mismatch between the physical bitrates used for data and ACK frames. Note according to the 802.11 specification the ACK frame is sent on a much lower base rate. However, when using the same bitrate for data and ACK frame the RMSE drops to around 3 percentage points (yellow). Unfortunately, adapting the bitrate used for transmission of ACK frames would require modifications to the STAs as well.

\medskip
\noindent \textit{\textbf{Takeaways: }}The RMSE for the UL is about 9 and 3 percentage points for standard and proposed ACK rate adaptation, respectively.

\iftoggle{techreport}{	
}%
{
\begin{figure}[h!]
	\centering
	\includegraphics[width=0.9\linewidth]{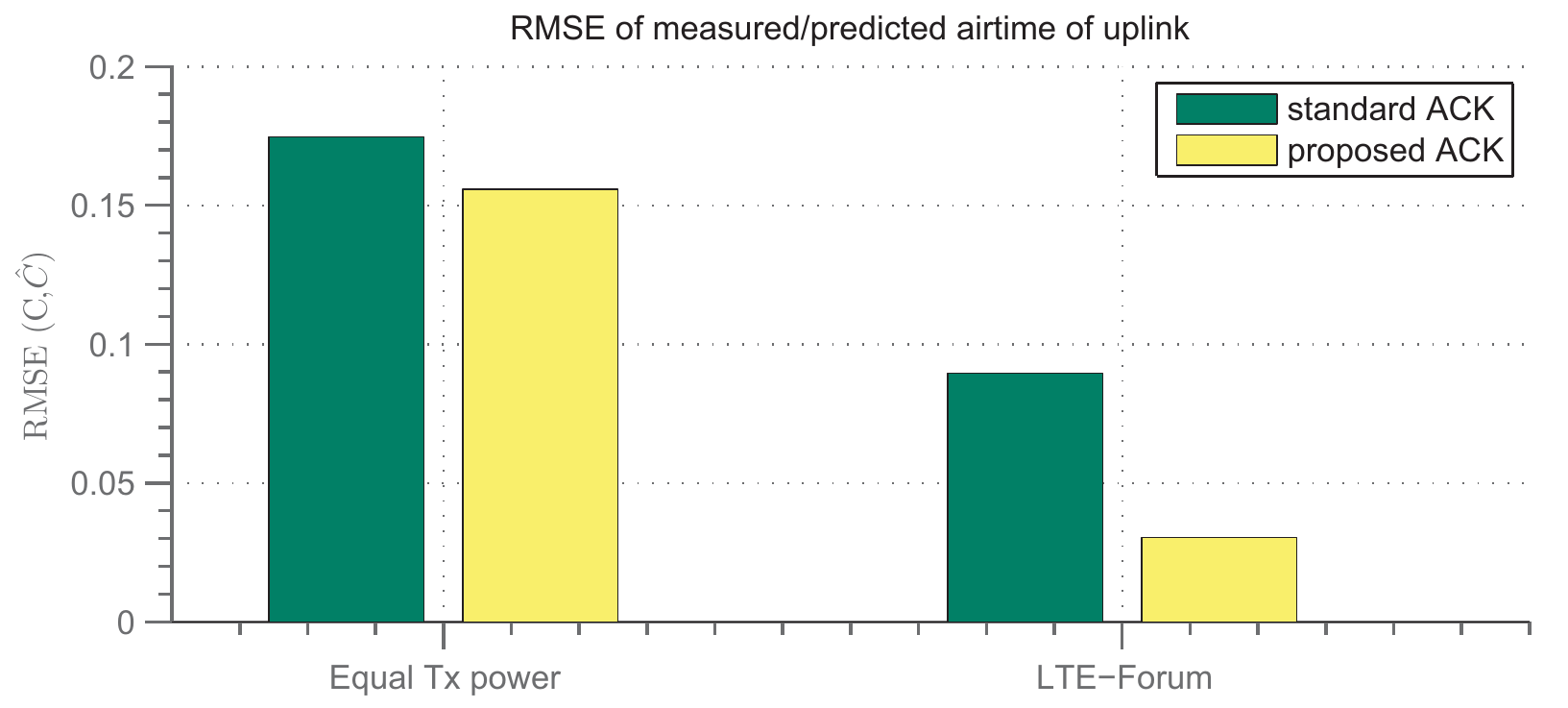}
	\caption{RMSE for uplink.}
	\label{fig:wiplus_ul_link_rmse_1_4}
\end{figure}
}

%% file: sections/related.tex
%
%
\section{Related Work}

WiPLUS is related to past work on LTE-U/WiFi co-existence schemes~\cite{zhang-2015, sagari-2015, almeida-2013, Al-Dulaimi-2015, zhang-2015-2, ko-2016} that either require modifications of WiFi STAs or LTE-U BSs. To the best of our knowledge, there is currently no solution which can be deployed on commodity hardware requiring only additional software modules at WiFi APs.

\iftoggle{techreport}{
	\begin{wrapfigure}[13]{R}{0.6\textwidth}
		\centering
		\includegraphics[width=1\linewidth]{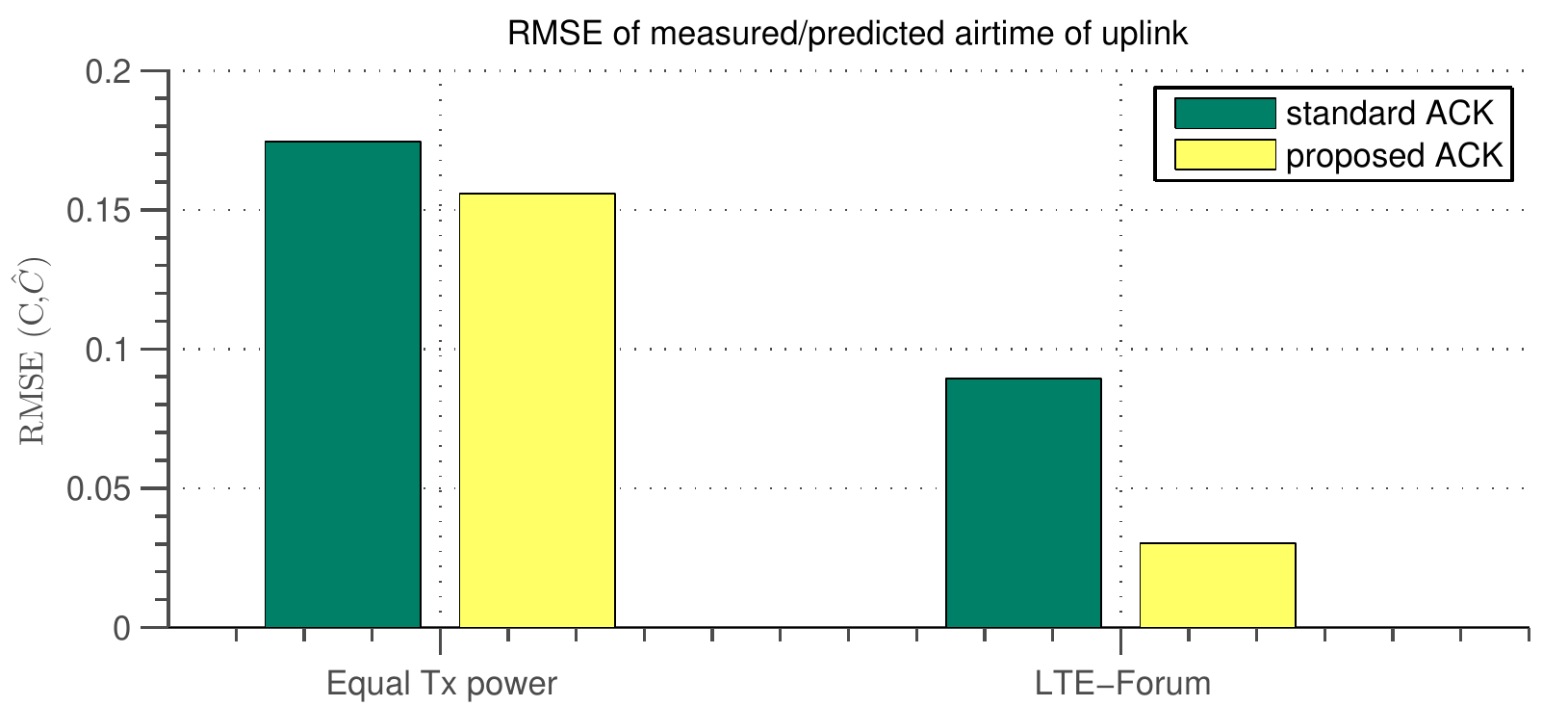}
		\caption{RMSE for uplink.}
		\label{fig:wiplus_ul_link_rmse_1_4}
		
	\end{wrapfigure}	
}%
{
}

Beside co-existence, there is a focus on detection of non-WiFi interference like Bluetooth or analog phones. The detection is done using spectrum analyzer functionality integrated in either custom~\cite{wispy, airsleuth} or commodity WiFi hardware~\cite{rayanchu2011airshark, rayanchu2012catching}. For instance, Airshark~\cite{rayanchu2011airshark} uses the spectral scanning capabilities of Atheros WiFi chips as input for classification of non-WiFi devices. Airshark is analyzing spectral scans at PHY layer, i.e. signal power received in each of the sub-carriers of an 802.11 channel. The drawback of incorporating only PHY layer information, is that it only allows the detection of interference within interference regime one (cf. Sec.~\ref{sec:under}). However, to enable the detection within the second interference regime, additional information from higher layers such as the MAC layer is needed, e.g. number of MAC retransmissions.

WiSlow~\cite{kim2014wislow} relies on measuring MAC layer information such as packet retries and correlates this information with the currently utilized PHY layer transmission rate. Therefore, WiSlow runs a packet capturer in the background. WiSlow allows identifying different sources of interference like microwave ovens for which it applies a duty cycle detection method that collects information about received ACK frames. While WiSlow can be extended to support the detection of LTE-U interference, its complexity stands in stark contrast to WiPLUS which only involves low-complexity operations, i.e. sampling of WiFi chip registers vs. extensive processing and inspection of each packet at line-speed.



%% file: sections/conclusion.tex
%
%
\section{Conclusions \& Future Work}

In this paper we introduced WiPLUS, a system that detects interfering LTE-U signals, computes their duty-cycles and derives the effective available medium airtime for each WiFi link in a WiFi BSS. WiPlus does not require any modifications to the WiFi client STAs and works with commodity WiFi cards installed in WiFi APs. For future work, we plan to utilize the HCCA functionality of 802.11e enabling explicit UL scheduling of STA traffic being interfered by LTE-U. 

\smallskip
\noindent \textbf{Acknowledgment:} The research leading to these results has received funding from the European Horizon 2020 Programme under grant agreement n688116 (eWINE project).